\documentclass[11pt, a4paper, twocolumn]{article}
\usepackage[utf8]{inputenc}
\usepackage[hmargin=1.5cm,vmargin=1.5cm]{geometry}

\usepackage{graphicx}
\usepackage{authblk}

\usepackage{url}
\usepackage{dcolumn}
\usepackage{amsmath}
\usepackage{amssymb}

\usepackage{hepunits}
\usepackage{xcolor}

\RequirePackage{fix-cm}
\RequirePackage{graphicx}

\newcommand{\lappdtm}{LAPPD\textsuperscript{TM}}
\newcommand{\pp}[1]{\left(#1\right)}
\newcommand{\isotope}[2]{$^{#2}$#1}
\newcommand{\tL}{t_{\rm{L}}}
\newcommand{\tR}{t_{\rm{R}}}
\newcommand{\tLAPPD}{t_{\rm{LAPPD}}}
\newcommand{\tPMT}{t_{\rm{PMT}}}
\newcommand{\tTrig}{t_{\rm{trig}}}
\newcommand{\dtLAPPD}{\Delta\tLAPPD}
\newcommand{\dtPMT}{\Delta\tPMT}
\newcommand{\dt}{\Delta t}
\newcommand{\tauR}{\tau_{R}}
\newcommand{\ndf}{\text{ndf}}
\newcommand{\tz}{t_{0}}
\newcommand{\tzj}{t_{0j}}
\newcommand{\fC}{f_{\rm{C}}}

\newcommand{\fDark}{f_{\rm{Dark}}}
\newcommand{\tf}{t_{f}}

\newcommand{\fCj}{f_{\rm{C}}^{\pp{j}}}

\newcommand{\fDarkj}{f_{\rm{Dark}}^{\pp{j}}}
\newcommand{\sigmaj}{\sigma^{\pp{j}}}

\newcommand{\differential}[1]{\text{d}#1}
\newcommand{\plusminus}[2]{^{\,+#1}_{-#2}}
\newcommand{\sLAPPD}{\sigma^{\pp{\rm{LAPPD}}}}
\newcommand{\sPMT}{\sigma^{\pp{\rm{PMT}}}}

\newcommand{\theia}{\textsc{Theia}}

\begin{document}

\title{Cherenkov and Scintillation Separation in Water-Based Liquid Scintillator using an \lappdtm}

\author[1,2]{T. Kaptanoglu \thanks{tannerbk@berkeley.edu}}
\author[1,2]{E.J. Callaghan \thanks{ejc3@berkeley.edu}}
\author[3]{M. Yeh}
\author[1,2]{G.D. Orebi Gann}

\affil[1]{University of California, Berkeley, CA 94720-7300, USA}
\affil[2]{Lawrence Berkeley National Laboratory, Berkeley, CA 94720-8153}
\affil[3]{Brookhaven National Laboratory, Upton, NY 11973-500, USA}

\maketitle

\begin{abstract}
This manuscript describes measurements of water-based liquid scintillators (WbLS), demonstrating separation of the Cherenkov and scintillation components using the fast timing response of a Large Area Picosecond Photodector (LAPPD). Additionally, the time profiles of three WbLS mixtures, defined by the relative fractions of scintillating compound, are characterized, with improved sensitivity to the scintillator rise-time. The measurements were made using both an LAPPD and a conventional photomultiplier tube (PMT).

All samples were measured with an effective resolution $O\pp{100~\text{ps}}$, which allows for the separation of Cherenkov and scintillation light (henceforth C/S separation) by selecting on the arrival time of the photons alone. The Cherenkov purity of the selected photons is greater than 60\% in all cases, with greater than 80\% achieved for a sample containing 1\% scintillator. This is the first demonstration of the power of synthesizing low light yield scintillators, of which WbLS is the canonical example, with fast photodetectors, of which LAPPDs are an emerging leader, and has direct implication for future mid- and large-scale detectors, such as \theia{}, ANNIE, and AIT/NEO.
\end{abstract}

\section{Introduction}\label{sec:intro}

Historically, both water and liquid scintillators have played key technological roles as target materials in particle physics experiments, most notably in large, monolithic neutrino detectors \cite{IMB,Ahn:2006zza,Ahmad:2002jz,Fukuda:1998mi,kamland,Seo:2019shs,An:2016ses,Abe:2019vii,Adamson:2017gxd,IceCube:2018cha,Aguilar-Arevalo:2018gpe,Agostini:2020mfq,Andringa:2015tza}. The design of future detectors will build on this experience by combining the advantages of both technologies -- nominally, the high light yield and low energy threshold offered by liquid scintillators, and the direction-reconstruction capabilities that have been demonstrated using Cherenkov light from water. Such hybrid detectors will leverage the simultaneous detection of both scintillation and Cherenkov light to achieve enhanced reconstruction and particle identification (PID), leading to improved background discrimination. Current work in Borexino \cite{Agostini:2021bxc}, a conventional liquid scintillator detector, has demonstrated limited directional reconstruction using early photons, which are relatively Cherenkov-rich, and work is ongoing in several collaborations to improve on this,  e.g. \cite{snoplus_angular,Jinping:2016iiq}. Such efforts are fundamentally hindered by the dominance of the scintillation yield over the relatively few detectable Cherenkov photons. WbLS \cite{YehWbLS}, which is defined by the suspension of liquid scintillator in water, offers a reduced scintillation yield, which allows for improved selection of Cherenkov light. Hybrid materials of this kind can lead to advances in the detection of low energy solar neutrinos \cite{Bonventre:2018hyd}, the reduction of the solar neutrino background for neutrinoless double beta decay ($0\nu\beta\beta$) searches \cite{Elagin:2016zgp,Aberle:2013jba,Biller:2013wua}, as well as long-baseline and atmospheric neutrino physics, for example through robust $\pi^{0}$-tagging via improved PID. Antineutrino detection will benefit from enhanced background rejection and lower detection thresholds, with further gains possible from loading with an isotope \cite{WbLStalk} with more favorable neutron capture characteristics, e.g. gadolinium \cite{Beacom:2003nk}.

In addition to WbLS, hybrid detectors can utilize modern photodetectors with improved timing resolution to maximize separation between the prompt Cherenkov photons and the delayed scintillation light, as well as chromatic separation via arrays of dichroic filters \cite{Kaptanoglu:2018sus}. The former can be achieved using small PMTs; large-area PMTs, however, typically have a transit-time-spread (TTS) of, at best, 650~ps ($\sigma$) \cite{Kaptanoglu:2017jxo}. LAPPDs offer significantly higher timing resolution \cite{LAPPD:2016yng,Anderson09thedevelopment,testbeam,Angelico:2020xzt} with large sensitive areas, and thus can be used to achieve improved C/S separation.

This work is the first demonstration of the synthesis of two key technologies contributing to future hybrid detectors: WbLS and LAPPDs. Using WbLS as target material, we demonstrate C/S separation in a prompt time window using the fast timing of an LAPPD. Simultaneously, the emission time profile \cite{Onken:2020pnv,Caravaca:2020lfs} of the scintillation light is measured, for the first time using the enhanced timing resolution offered by an LAPPD. The measurements were performed using an $^{90}$Y source with Q-value of 2.28~MeV, relevant for future $0\nu\beta\beta$ and solar neutrino experiments. Similar measurements using the dichroicon \cite{Kaptanoglu:2019gtg}, a Winston cone concentrator built from dichroic filters, are anticipated.

This work takes place in the context of ongoing efforts to incorporate both WbLS and fast photodetectors into upcoming detectors. The experimental apparatus utilized the same electronics, data acquisition (DAQ) software, and muon veto panels from the CHESS setup used in previous work \cite{Caravaca:2020lfs,Caravaca:2016ryf}. Deployment of an LAPPD coincident with the CHESS array is under consideration. At a larger scale, the NuDOT detector, an upgrade to FlatDOT \cite{Gruszko:2018gzr}, will employ novel scintillation mixtures to achieve C/S separation using fast timing and advanced reconstruction techniques, in a tonne-scale detector \cite{nudot}. Most notably, the ANNIE collaboration plans to deploy both WbLS and LAPPDs to aid in vertex reconstruction and improve neutron detection efficiency \cite{Back:2017kfo}. Future detectors, such as AIT-NEO \cite{Askins:2015bmb} and \theia{} \cite{Askins:2019oqj}, may also utilize WbLS and LAPPDs, as well as photon sorting devices, such as dichroicons. The measurements presented here constitute the first combination of multiple of these technologies, and provide insight for their incorporation into larger apparatus.

\section{LAPPDs}\label{sec:lappd}

LAPPDs are photodetectors that utilize stacked microchannel plates (MCPs) and a photocathode in a planar geometry, vacuum sealed in a glass or ceramic body. The development and features of LAPPDs are described in detail in Refs. \cite{Lyashenko:2019tdj,osti_1564252,Malace:2021hma}. We give here a high-level overview and describe the specific details relevant for the measurements performed in this work.

\begin{figure}[t!]
    \centering
    \includegraphics[clip=true, trim=15cm 0cm 15cm 28cm, width=0.40\textwidth]{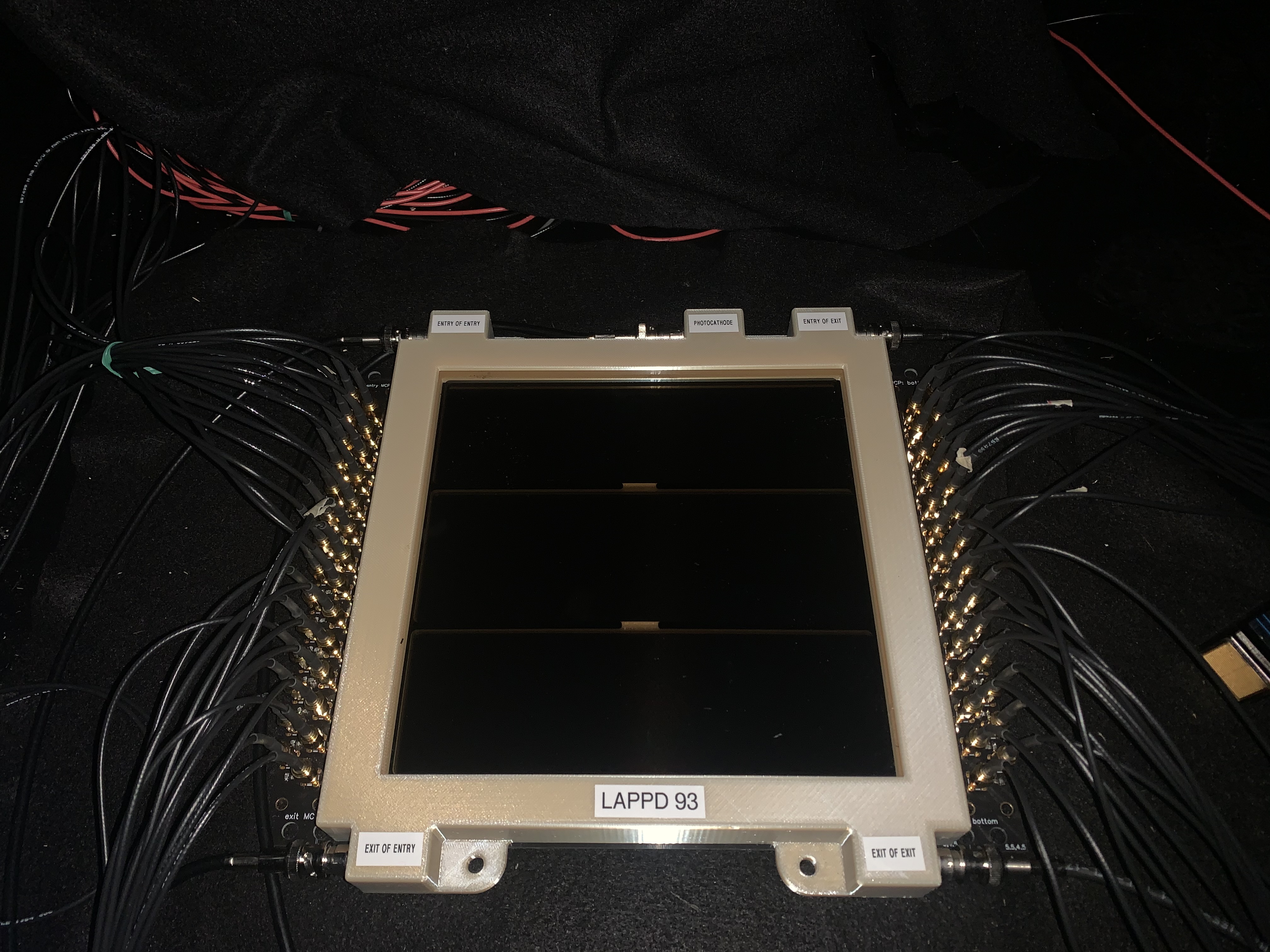}
    \caption{Tile \#93, fully cabled for the readout of 28 strips and to provide high voltage to the four MCPs and the photocathode. The sensitive area is about 380~cm$^{2}$.}
    \label{fig:lappd}
\end{figure}

LAPPD tile \#93 was produced by Incom Inc. (sealed on 2/3/2021) and is shown in Figure \ref{fig:lappd}. The tile was acquired in March of 2021 and is used for the measurements in this paper. Photons incident on the photocathode release photoelectrons (PEs) that are multiplied by the MCPs. The resulting electron cloud is collected by a strip-line anode with twenty-eight silver strips. Direct signal read out at both ends of each strip (“left- and right-hand sides") allows for enhanced timing resolution across the full dimension of the device, as well as the reconstruction of the intrastrip position of the photon (Incom also offers a pixelated version of the LAPPD that features indirect capacitively coupled readout of signals detected on a printed circuit board coupled to the bottom of the tile \cite{incom_communication}). The LAPPD window is 5mm thick and is made of fused silica glass with a Multi-Alkali (K2NaSb) photocathode deposited on the inside surface; the tile has a sensitive area of approximately 380~cm$^{2}$. There are two narrow spacers that run parallel to the strips, required for structural support, which create small dead areas. An alternative design employs X-shaped spacers \cite{Lyashenko:2019tdj}; the horizontal spacers used here leave more of the total area sensitive, including the center of the tile.

\begin{figure}[b!]
        \centering
        \includegraphics[width=0.40\textwidth]{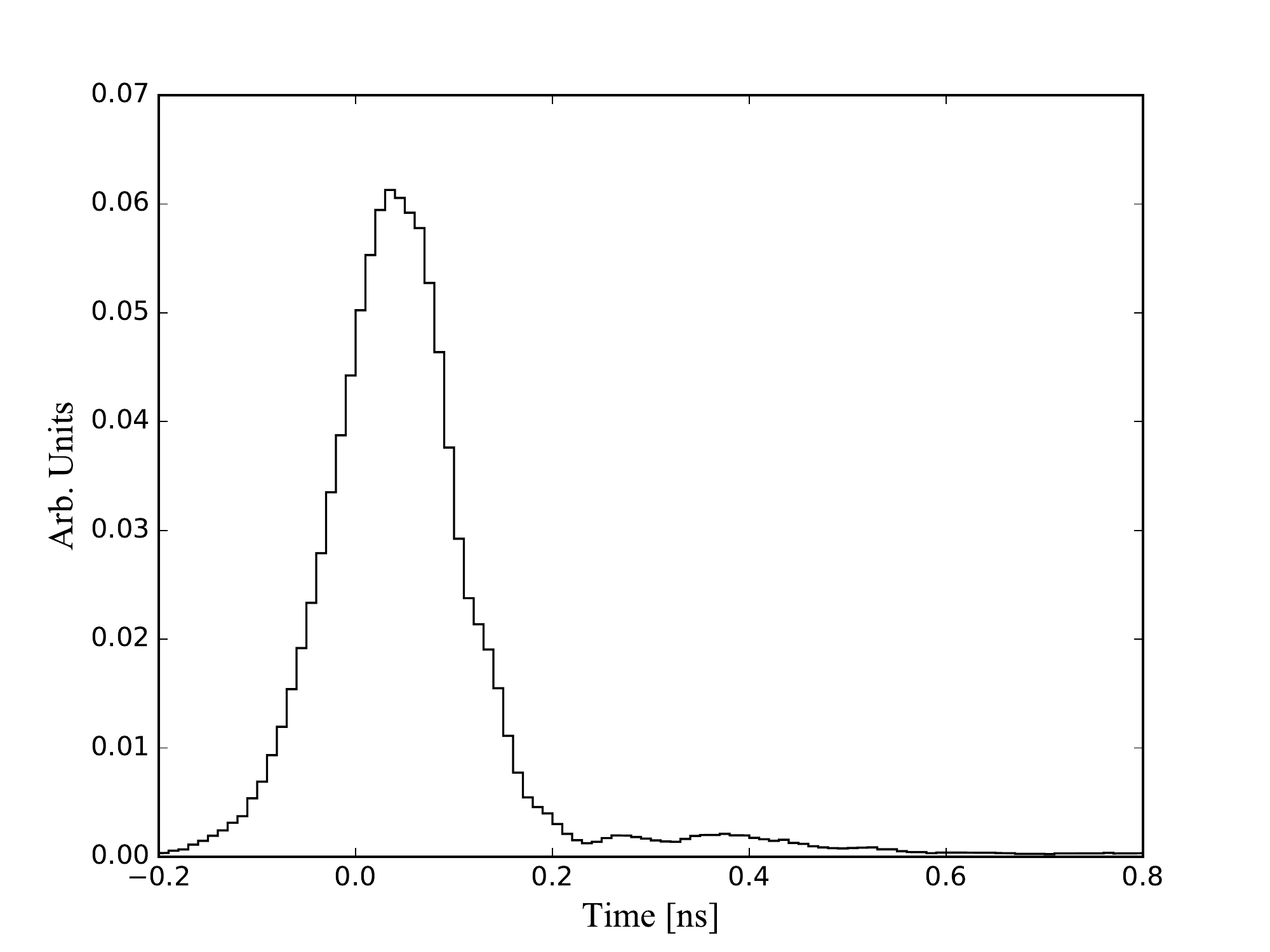}
        \caption{The LAPPD transit time distribution as measured and provided by Incom Inc. \cite{incom_communication}. The TTS is about 70~ps. The distribution has been smoothed for visualization purposes.}
        \label{fig:lappd_tts}
    \end{figure}

Characterizations of previous LAPPDs have been performed \cite{Lyashenko:2019tdj,Jocher:2018yhc}, and show single photon sensitivity with high gain (on the order of 10$^{7}$), timing resolution below 100~ps, reasonably high quantum efficiency (QE) (LAPPD tile \#93 has an average QE at 365 nm of 28.3\% and a maximum QE of 31.3\% \cite{incom_communication}), millimeter-scale spatial resolutions, and dark rates around 100~Hz/cm$^{2}$. The dark rate of tile \#93 is atypically high, more than an order-of-magnitude above this average \cite{incom_communication}. The critical feature of the LAPPD in the present work is its timing resolution, which for previous tiles has been measured to be less than 80~ps \cite{Lyashenko:2019tdj}. The transit time distribution of tile \#93 is shown in Figure \ref{fig:lappd_tts}, measured by Incom using a pulsed laser \cite{incom_communication}. The observed TTS is approximately 70~ps ($\sigma$), but the measurement is limited by the laser and electronics of the system. The late pulses arriving after the prompt peak are caused by scattering of the photoelectrons inside of the LAPPD.
    
The enhanced timing and spatial resolution of LAPPDs make them a candidate technology for use in future optical detectors. Fast-timing photodetectors offer a route to achieving C/S separation in large detectors, and LAPPDs acting in this role may contribute to reaching this goal. Additionally, both the improved timing and pixelation of LAPPDs lead to enhanced vertex reconstruction in larger detectors. This leads to improved PID, such as distinguishing $e^{-}$ events from $e^{+}$, $\gamma$, or $\alpha$ particles \cite{Dunger:2019dfo}. Studies for the ANNIE detector have shown that the introduction of five LAPPDs (to a baseline design of 125 PMTs) enhances vertex resolution from about 40 to 20~cm for $\mu^{-}$ events above 200~MeV \cite{Tiras:2019ozv}. Similarly, simulations of a 200~kton water detector show that employing sensors with LAPPD-scale timing leads to significant improvements in vertex resolution. The improvements provided by LAPPDs to position and direction reconstruction in large hybrid detectors, at scales ranging from one to fifty kilotonnes, and the consequences for sensitivity of $0\nu\beta\beta$ and precision CNO measurements, are studied in Ref. \cite{Land:2020oiz}.

\section{Target Materials}\label{sec:materials}

    This work considers three samples of WbLS, formed by combining linear alkylbenzene (LAB) and 2,5 diphenyloxazole (PPO), a common solvent-fluor pair in neutrino detectors \cite{Abusleme:2020bbm,Anderson:2020xxb}, with water. It is a candidate target material for upcoming optical detectors, including ANNIE \cite{Back:2017kfo}, AIT-NEO \cite{Askins:2015bmb}, and \theia{} \cite{Askins:2019oqj}. This work considers WbLS mixtures prepared with scintillator loaded at the 1\%, 5\%, and 10\% levels.

WbLS offers several advantages over pure liquid scintillator. Having a high water concentration, the absorption length is similar to pure water, which is important in large, monolothic detectors, where the average path length of detected photons may be many meters. The scintillation light yield is reduced relative to a pure LAB + PPO mixture, which enhances the Cherenkov signal amidst the relatively large amount of scintillation light. Additionally, the fraction of scintillator within the mixture is controllable, giving WbLS the unique characteristic that the purity of the Cherenkov light selection can be balanced against a higher light yield. In other words, it is possible to determine the WbLS mixture that will optimize sensitivity for the physics goals of an experiment prior to deploying that mixture in the detector. Additionally, WbLS is less expensive, per unit volume, than pure scintillator -- an advantageous feature in the next-generation regime of detectors at the scale of tens of kilotons.

The total scintillation light yield of several WbLS mixtures has been characterized using the CHESS setup \cite{Caravaca:2020lfs}, which showed that the light yield scales roughly linearly with the level of scintillator loading. The scintillation emission spectrum of the WbLS mixtures was measured in \cite{Onken:2020pnv} using an X-ray source. The scintillation time profile has been characterized using both a $\beta$ source \cite{Caravaca:2020lfs} and X-ray excitation \cite{Onken:2020pnv}, which demonstrate that the emission timing is faster than LAB + 2~g/L PPO. This makes C/S separation more challenging, as more scintillation light is emitted promptly, coincident with the Cherenkov component. We demonstrate that such separation is still achievable using fast timing photodetectors.

Pure liquid scintillators are often deoxygenated by sparging with an inert gas, which mitigates quenching effects due to the presence of dissolved atmospheric oxygen -- for example, the time profile of LAB + PPO cocktails is dependent on the oxygen concentration \cite{OKeeffe:2011dex}. In WbLS, however, such quenching is likely dominated by the abundant water molecules, with dissolved oxygen playing a smaller role. Indeed, measurements made using the methodology described in this paper did not exhibit any significant change to the time profile after sufficient nitrogen sparging. The results reported in this work are for samples that have been exposed to the atmosphere.

\section{Experimental Setup}\label{sec:setup}

\begin{figure}[t!]
    \centering
    \includegraphics[width=0.45\textwidth]{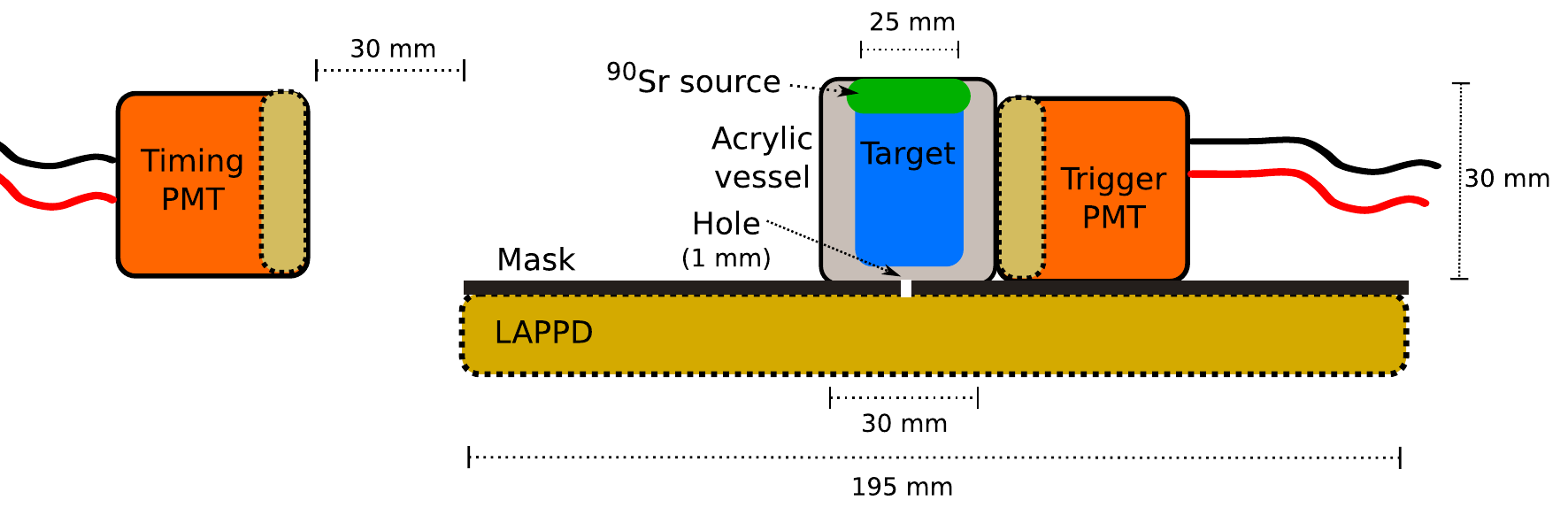}
    \caption{The experimental setup used for the measurements in this paper. A $^{90}$Sr $\beta^{-}$ source is deployed above a liquid target contained in a cylindrical UVT acrylic vessel, above an LAPPD which is used to detect single photons, which are mostly blocked by a black mask. A PMT located $\sim$25~cm away simultaneously detects single photons. An identical PMT is optically coupled to the acrylic vessel and used as trigger.}
    \label{fig:setup_drawing}
\end{figure}

The experimental setup, shown in Figure \ref{fig:setup_drawing}, utilizes a \isotope{Sr}{90} button source, purchased from Spectrum Techniques \cite{spectrum_techniques}, placed in a cylindrical, UV-transparent acrylic vessel filled with each target liquid. The \isotope{Sr}{90} $\beta$-decays with Q-value 546~keV to \isotope{Y}{90}, which then $\beta^{-}$ decays with Q-value 2.28~MeV and half-life 64 hours. The vessel is 30~mm in both diameter and height, with the target material occupying an inner cylindrical volume of diameter 20~mm. The source is placed directly above the target material and rests on a ledge that is 3.2~mm thick. There is no acrylic separating the source container and target material, which maximizes the fraction of $\beta$ energy deposited into the scintillator.

The acrylic vessel is placed on top of the LAPPD, which is optically masked off, except for a circular hole 1~mm in diameter located near the center of the LAPPD. The vessel is centered on the hole and is optically coupled to the LAPPD using Eljen Technology EJ-550 optical grease \cite{eljin}. The hole is approximately centered on strip 14, which is the only strip utilized in the analysis. The use of the mask is crucial to the measurement as it ensures that the LAPPD operates in the single photoelectron (SPE) regime across all samples, an important requirement of the coincidence technique employed, first described in Ref. \cite{osti_1564252}. The diameter of the hole was selected to ensure SPE operation, which is confirmed both by the low coincidence rate and through Monte Carlo (MC) simulations of the setup. The LAPPD is situated horizontally on top of a sealed plastic scintillator panel, which is used to reject events in which downward-going muons traverse the body of the LAPPD.

A 1-inch square Hamamatsu PMT (model H11934-200) \cite{r11934} is optically coupled to the side of the vessel and used to trigger the DAQ, and correct all measured photon times for arbitrary delays associated with the triggering logic. The trigger threshold is set to 15~mV, corresponding to 3-4~PE.

A second H11934-200 PMT, referred to as the timing PMT, was placed $\sim$25~cm away from the center of the source. The relatively high noise rate of LAPPD \# 93 limits its sensitivity to the long time-scale behavior of the scintillation time profile; including the PMT in the measurement constrains the scintillation time profile at longer times, and mitigates the degeneracy between the scintillator rise-time and an overall system timing delay.

The LAPPD is powered such that there are 200~V biases between the photocathode and first MCP, across the gap between the two MCPs, and between the anode and the bottom MCP, and 850~V biases across each MCP. The net bias across the device is 2300~V. The trigger and timing PMTs are each biased to 950~V. All signals are digitized using a CAEN V1742 digitizer \cite{caen} over a 1~V dynamic range, sampling at 5 GHz for 1024 samples, yielding waveforms that are 204.8~ns in length. The data is read out over USB and custom DAQ software is produces \texttt{HDF5} files contain the raw waveforms.

\section{Waveform Analysis}\label{sec:waveform_analysis}

The digitized waveforms are analyzed using custom analysis code which reads the \texttt{HDF5} files and performs pulse processing on the component waveforms. An identical analysis chain is applied to the data and MC.

The DAQ is triggered such that prompt light on each channel arrives no earlier than 30 ns after the beginning of the digitization window. A per-waveform baseline is calculated using a 15 ns window preceding the arrival of prompt light, which is critical to robustly measure the amplitude of each pulse. Events with unstable baselines, usually due to one-off readout errors specific to the employed digitizers, are rejected from analysis. The sacrifice of this cut is approximately 0.2\%.

Single photon pulses are found by applying 5~mV threshold across a 160~ns window in the LAPPD and timing PMT waveforms, chosen to minimize crossings due to electronic noise while accepting a majority of SPE pulses. A 15~mV threshold is applied to the trigger PMT waveforms. Events in which any of the waveforms cross threshold more than once, which are caused by pickup from electronics elsewhere or the detection of multiple photons, are rejected. A 10~mV threshold is applied to the PMTs coupled to the muon panel below the LAPPD to identify and reject muons, which can rapidly ionize the MCPs, generating a large signal ultimately leading to pickup on virtually all channels.

The timing associated with an SPE-like pulse is determined by applying constant-fraction discrimination (CFD) to the waveform, linearly interpolating between samples when necessary. The fractional threshold is arbitrary, and in this work is set to 60\%. This procedure is applied to the signal from the timing PMT, producing a time value $\tPMT$, as well as the left- and right-hand channels associated with strip 14 to produce two values: $\tL$ and $\tR$. The times $\tL$ and $\tR$ contain anticorrelated contributions from the time for the signal created in the anode strip to propagate to either side of the device for readout, which is mitigated by defining a strip-level time value, $\tLAPPD$, as their average.

A simple threshold-crossing time is assigned to the trigger PMT, which is more robust to fluctuations in multi-PE pulse shape associated with the distribution of individual photon arrival times. The threshold used in this work is 3~mV, which is above fluctuations from electronic noise but will be crossed by the signal from the first detected photon. This value is $\tTrig$.

After correcting for arbitrary delays associated with the cabling and trigger logic, the ``hit time'' of a photon in the LAPPD is defined as:
\begin{equation}
    \dtLAPPD = \tLAPPD - \tTrig = \frac{\tL + \tR}{2} - \tTrig.
\end{equation}
Similarly, the hit time associated with a photon in the timing PMT is defined as:
\begin{equation}
    \dtPMT = \tPMT - \tTrig.
\end{equation}

The charge collected in the trigger PMT can be used as a proxy for the number of photons detected, itself used as proxy for the energy deposited in the sample. The charge is calculated by integrating the waveform in a dynamic window extending from 14~ns before to 24~ns after its peak. In each dataset, events with charge less than 25\% of the spectrum endpoint are removed from consideration. This removes the \isotope{Sr}{90} events and leaves a pure sample of \isotope{Y}{90} decays with energy above $\sim$570~keV, as verified using the MC described in Section \ref{sec:simulation}.

\section{Simulations}\label{sec:simulation}

A detailed MC of the trigger PMT and LAPPD is implemented in \texttt{RAT-PAC} \cite{ratpac}, a \texttt{GEANT4}-based \cite{GEANT4:2002zbu} simulation package, to model the production, propagation, and detection of photons, using a realistic detector geometry and DAQ response. The focus is a detailed modeling of the LAPPD response, and as such the timing PMT is not included in the simulation.

The production and propagation of photons through the target material is determined using a GLG4Scint-based optical model. The optical properties of WbLS, used as input to the model, are either measured values or estimates computed from the properties of pure water and LAB-based liquid scintillator.  Several of the parameters which are expected to have significant impact on the results of this work, such as the emission spectrum \cite{Onken:2020pnv}, have been measured. The scintillation time profile is determined in this work, and the light yield values were measured in \cite{Caravaca:2020lfs}, which in this work are subject to additional scalings to account for uncertainties in the light yield, index of refraction, and calibration of the photodetectors. Other parameters are expected to have negligible impact on this work, given the small sample size involved. Further details can be found in \cite{Caravaca:2020lfs}.

\begin{figure}[t!]
    \centering
    \includegraphics[width=0.40\textwidth]{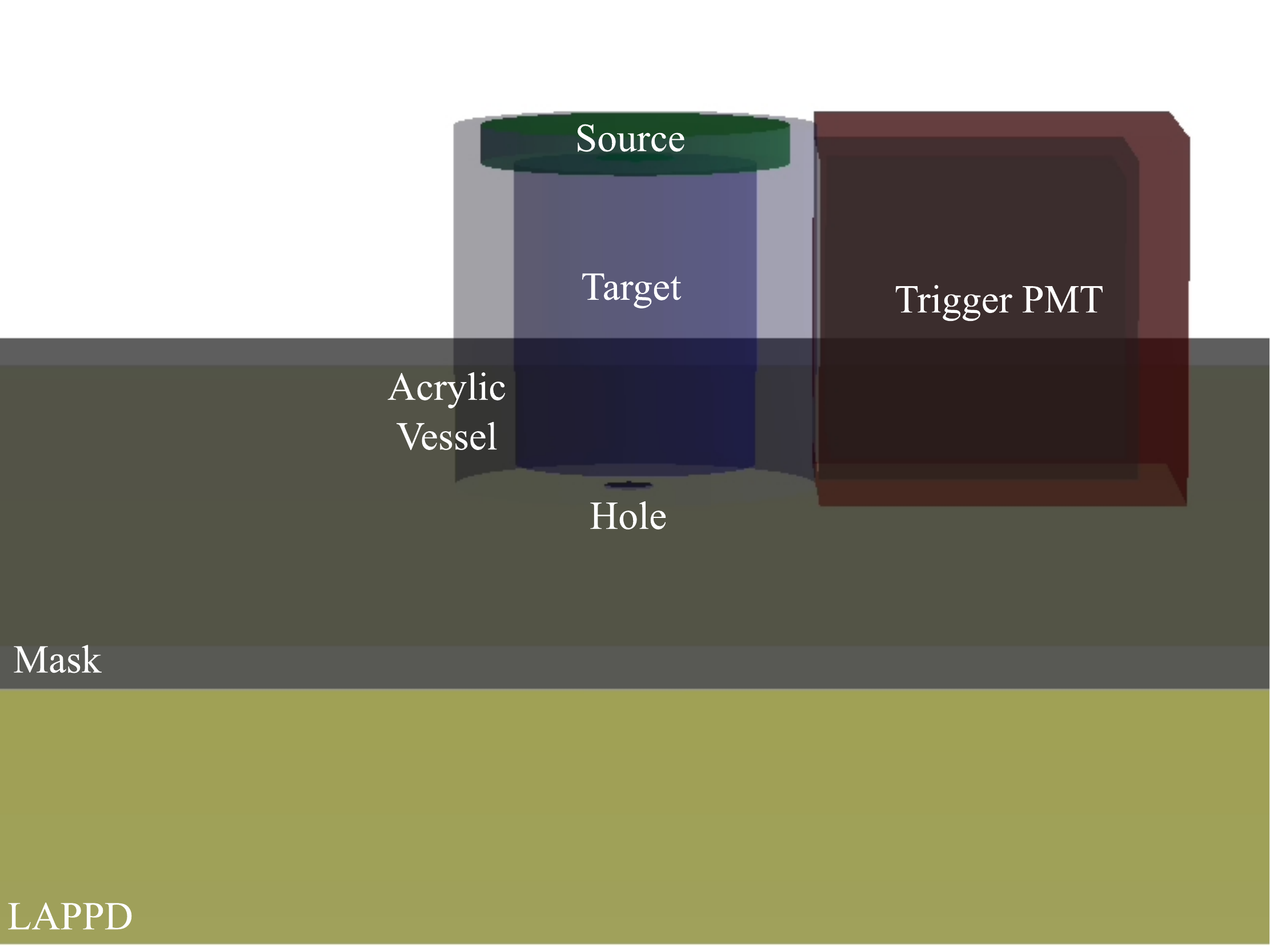}
    \caption{The geometry of the setup, shown in Figure \ref{fig:setup_drawing}, visualized looking from the side of the LAPPD in \texttt{RAT-PAC}.}
    \label{fig:geo_viz}
\end{figure}

The strip-segmented LAPPD is modeled as 28 distinct detectors, which neglects charge-sharing between neighboring strips --- an acceptable approximation in this single-strip measurement, which largely operates in the SPE regime. The nominal QE and transit time distributions were provided by Incom \cite{incom_communication}, measured using an UV LED and pulsed laser, respectively. No dark hits are simulated in the LAPPD; ideal comparisons between the data and MC are made by subtracting the dark rate from the data. The QE and TTS of the trigger PMT are taken from the Hamamatsu datasheet \cite{r11934}. The mask placed on top of the LAPPD is modeled as perfectly absorbing. The full geometry implemented in \texttt{RAT-PAC} is shown in Figure \ref{fig:geo_viz}.

The DAQ is modeled by generating analog pulses from all photodetectors, as detailed in Section \ref{sec:calibrations}, and applying trigger logic and realistic digitization. Multi-PE waveforms are constructed by linearly adding independent SPE pulses. The output of the MC are \texttt{HDF5} files identical in structure to the data.

\section{Calibrations}\label{sec:calibrations}

The shapes and sizes of the analog pulses produced by the photodetectors, as well their effective detection efficiencies and timing characteristics, must be calibrated in order to accurately model the measurement apparatus using the MC. An LED is used to calibrate single-photon pulses from the LAPPD and trigger PMT, and subsequently a Cherenkov-pure water dataset is used to tune their efficiencies and timing characteristics. These procedures are described in detail below.

\subsection{Pulse shape calibration}\label{sec:calibration_pulses}

Robustly matching the distributions of $\dt$ generated via MC to those observed in data requires accurate modeling of the pulses generated by the LAPPD and PMTs. Calibration of the pulse shapes is performed with a pure SPE dataset collected using a pulsed LED. For both the LAPPD and trigger PMT, the LED was arranged and powered such that the coincidence rate for a signal out of the photodetector was less than 1\%, which ensures that the devices operate exclusively in the SPE regime.

\begin{figure}[t!]
    \centering
    \includegraphics[width=0.45\textwidth]{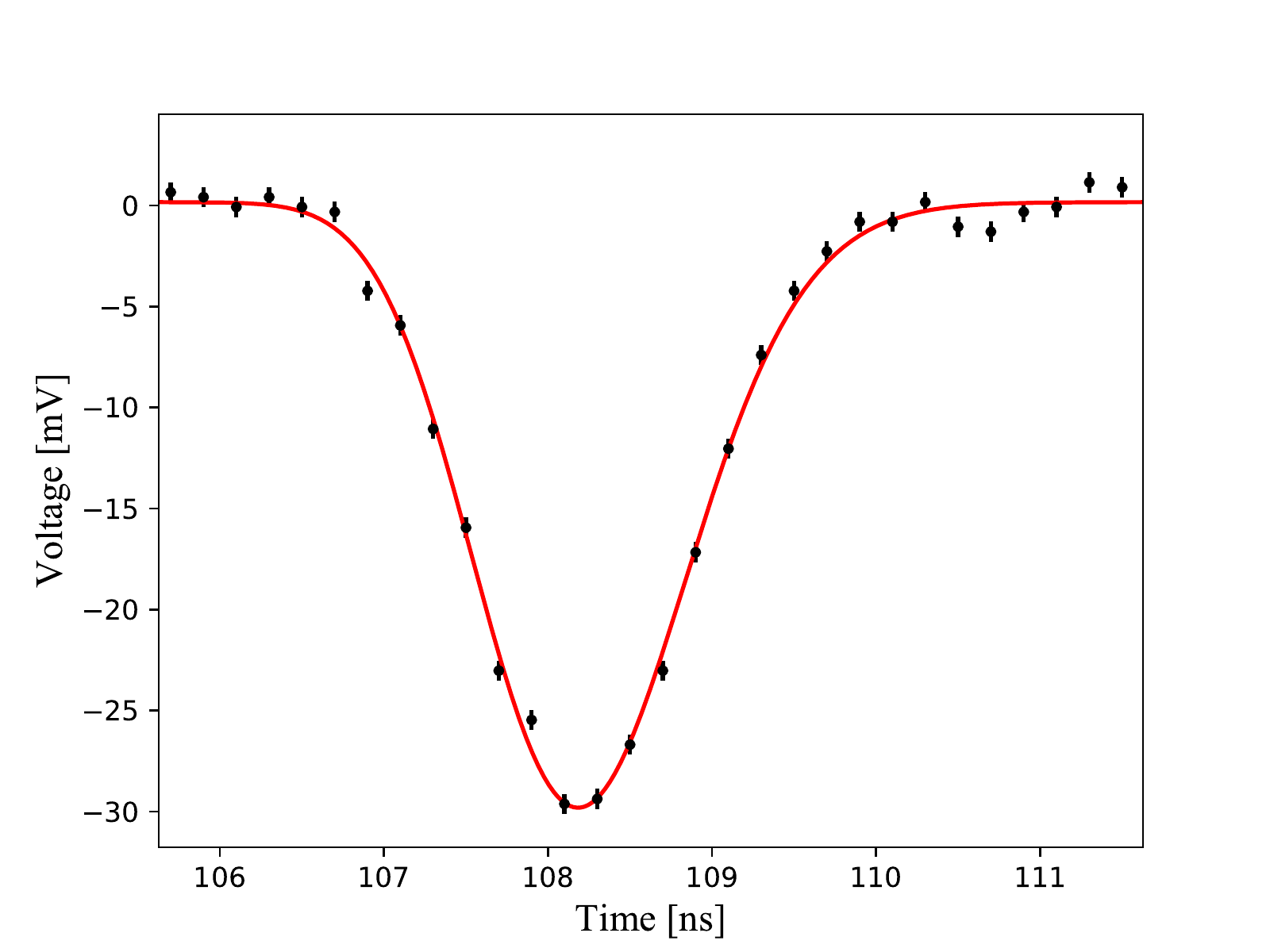}
    \caption{An example log-normal fit to an SPE waveform from the LAPPD.}
    \label{fig:lappd_waveform_fit}
\end{figure}

The digitized SPE-like pulses are individually fit with a log-normal function

\begin{equation}\label{eq:lognormal}
    f\pp{t} = B
            + \frac{Q}{(t - \tz)\sqrt{2\pi}\sigma}
              e^{-\frac{1}{2}\pp{\log\pp{\frac{t - \tz}{m}}/\sigma}^{2}},
\end{equation}
where $B$ is the baseline, $Q$ is the charge contained in the pulse, $\tz$ is arrival time of the pulse, and $m$ and $\sigma$ are shape parameters. An estimate of the electronic noise is taken as the standard deviation of a baseline window defined early in the waveform, and is used as an uncertainty on each voltage sample. The fit is then performed by $\chi^{2}$-minimization. An example fit to a waveform from the LAPPD is shown in Figure \ref{fig:lappd_waveform_fit}.

For each device, about 50,000 waveforms were collected and fit. Well-formed pulses were selected by requiring the minimal $\chi^{2}/\ndf < 2.5$. The LAPPD pulses from the left- and right-hand sides of the strip are fit separately and are generally consistent with one another. For each parameter, the fit results for both sides are combined into a single histogram. The histograms are fit with Gaussian distributions, which are then sampled from in the MC. The results of this calibration are shown in Table \ref{tab:calibration_parameters}. 

\begin{table}[b!]
    \small
    \centering
    \begin{tabular}{l r l l}
         \- &
            \textbf{Parameter} &
            \textbf{Mean} &
            \textbf{Std. Dev.} \\
         \hline \hline 
         \textbf{Strip 14} &
            $\sigma~\text{[ps]}$ &
            69 &
            1.0 \\
         \- &
            $m~\text{[ns]}$ &
            8.5 &
            0.05 \\
         \- &
            $Q~\text{[pC]}$ &
            0.41 &
            0.23 \\
         \hline
         \textbf{Trigger PMT} &
            $\sigma~\text{[ps]}$ &
            360 &
            50 \\
         \- &
            $m~\text{[ns]}$ &
            6.1 &
            0.25 \\
         \- &
            $Q~\text{[pC]}$ &
            0.75 &
            0.48 \\
    \end{tabular}
    \caption{Results of the pulse-shape calibration used as input to the MC, for waveforms from the LAPPD and trigger PMT.}
    \label{tab:calibration_parameters}
\end{table}

\subsection{Water Data}\label{sec:calibration_tts}

Due to variations in the nominal operating characteristics of the photodetectors, shortcomings of the optical model employed in MC, and potentially a breakdown of the linear pulse superposition assumption at the trigger PMT, na\"ive simulation of the setup using nominal input specifications will not yield a perfect model of the measured timing distribution. A water dataset, taken using the same geometry as described in Section \ref{sec:setup}, consists of pure Cherenkov light, the production of which is well understood, and as such provides a benchmark for tuning the simulation inputs to better model the observed data. In particular, we tune the QE and TTS of the trigger PMT, each of which has a direct impact on the observed timing distributions.

\begin{figure}[t!]
    \centering
    \includegraphics[width=0.45\textwidth]{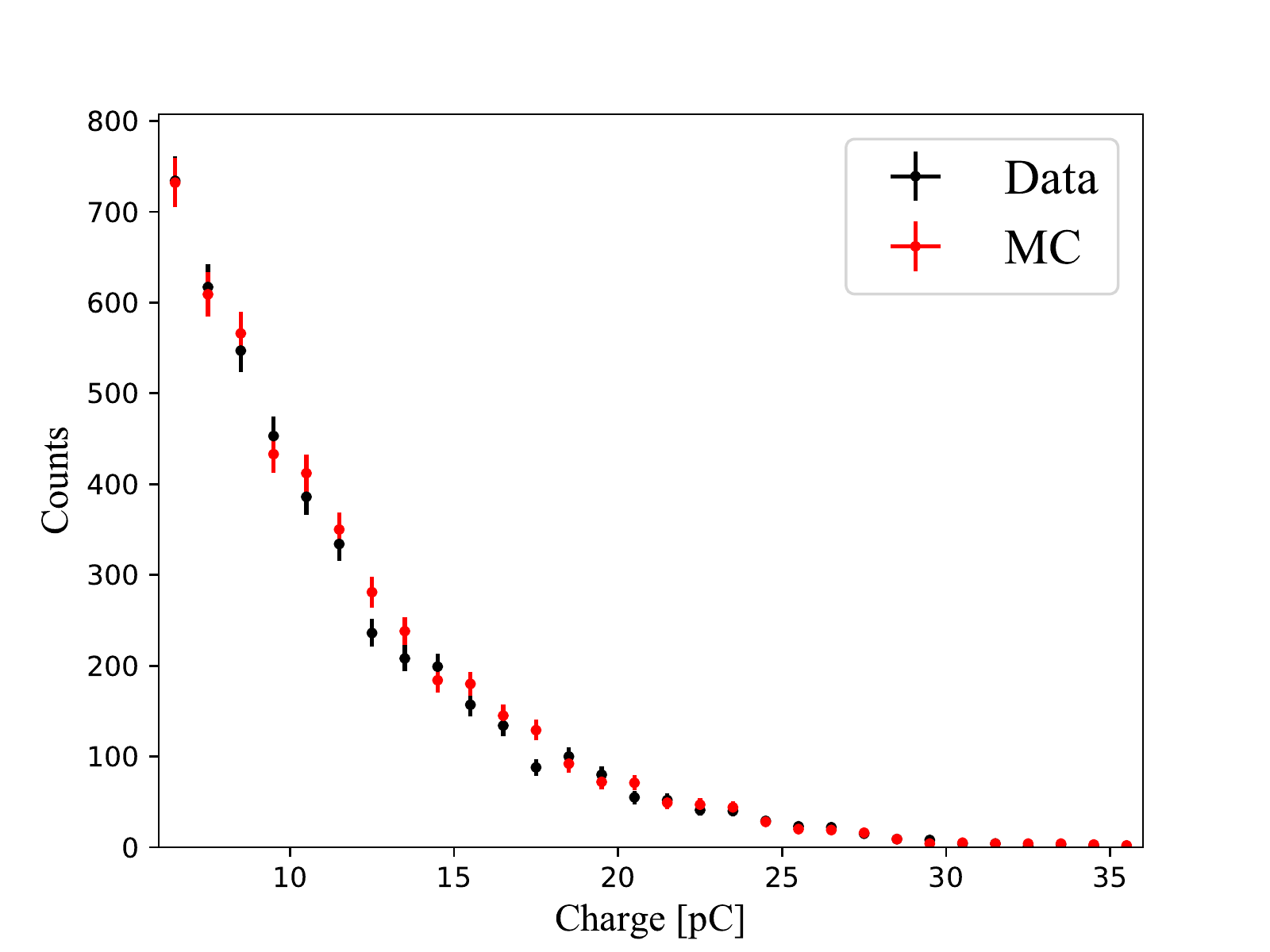}
    \caption{Efficiency-calibrated trigger PMT charge distribution, in both data and MC, for a water target. The $\chi^{2}$/\ndf{} comparing the data and MC is 22.9/30.}
    \label{fig:water_q_result}
\end{figure}

As described in Section \ref{sec:analysis}, the occupancy of the trigger PMT contributes to an effective timing resolution of the system, and as such must be properly modeled. To achieve this, we tune the QE of the trigger PMT by comparing the predicted and measured charge distributions as a function of the efficiency scaling. The MC was evaluated in efficiency steps of 1\%, and the minimum $\chi^{2}$ is observed at a scaling of 110\%. A comparison of the charge distributions in data and MC is shown in Figure \ref{fig:water_q_result}. A tuning at the level of 10\% scale is acceptable, as Hamamatsu only provides a typical QE curve, and not a dedicated measurement of the PMT utilized in this work.

The TTS of the trigger PMT is calibrated in a similar fashion: the $\dtLAPPD$ distribution of events with trigger charge greater than 6.0~pC (see Section \ref{sec:waveform_analysis}) is compared between data and MC, as a function of the input TTS. The timing distributions in data and MC are each uncorrected for global offsets associated with delays due to cable lengths and trigger logic, and are aligned for each comparison by minimizing the $\chi^{2}$ between them. The TTS is scanned in steps of 5~ps, yielding an optimal TTS of 185~ps ($\sigma$), compared to the nominal specification of 115~ps. This inflation may be due in part to deviations of the TTS of the photodetectors from their standard specifications, or due to mismodeling of some of the optics, such as the Rayleigh scattering within the acrylic vessel.

\begin{figure}[t!]
    \centering
    \includegraphics[width=0.45\textwidth]{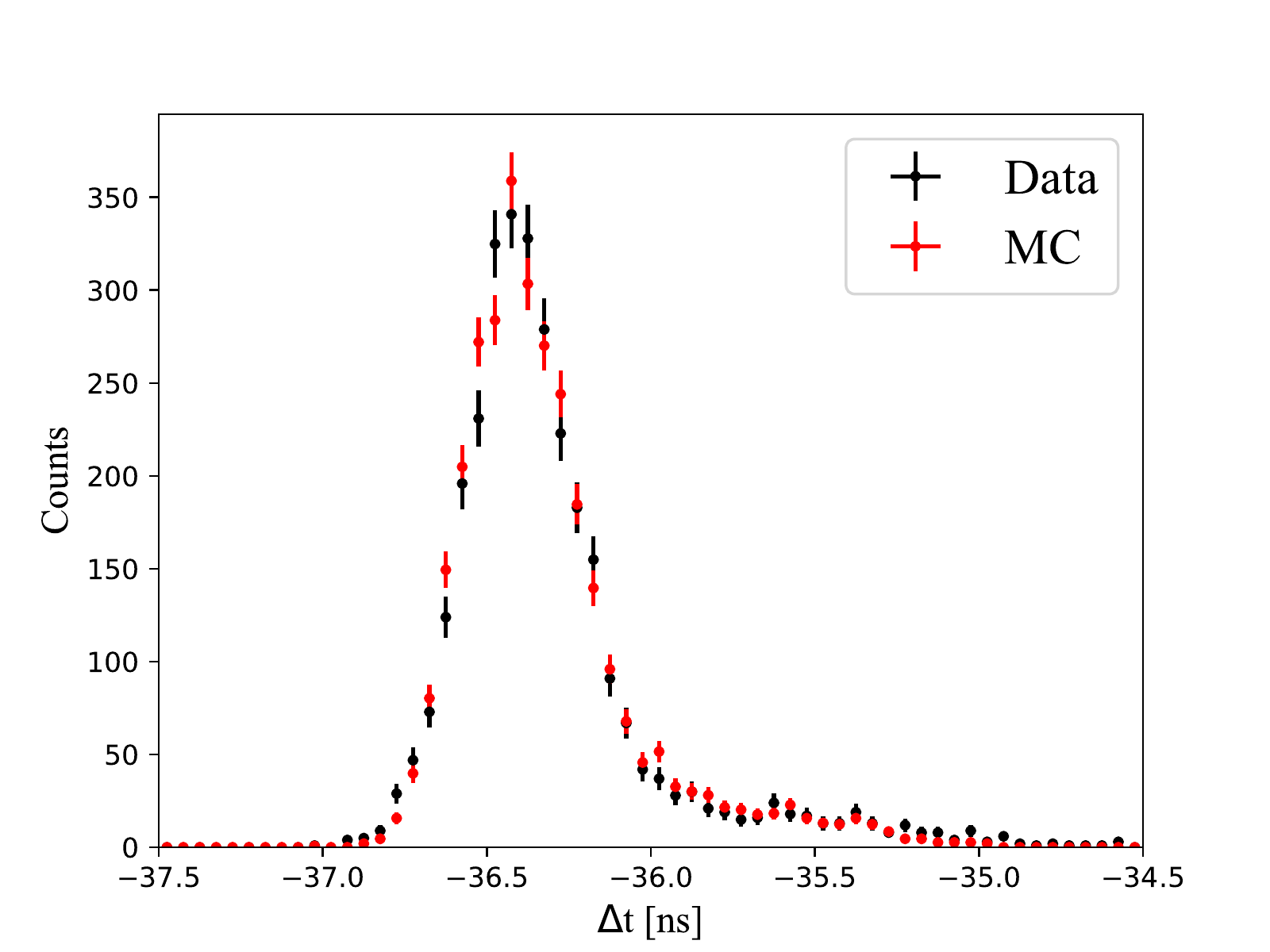}
    \caption{TTS-calibrated timing distributions, in both data and MC, for a water target. The $\chi^{2}$/\ndf{} comparing the data and MC is 41.9/40.}
    \label{fig:water_result}
\end{figure}

The calibrated timing distribution is shown in Figure \ref{fig:water_result}, which exhibits a dominant Gaussian component and a modest tail of late light. By fitting a Gaussian to a window extending $\pm$ 300~ps around the peak, the effective resolution of the water data is found to be approximately 164 $\pm$ 4~ps, a demonstration of the fast timing capability of the LAPPD. The tail of the timing distribution consists of photons which scatter along their trajectory from production to detection at the LAPPD, as well as from the backscatters of PEs within the LAPPD.

\section{Analysis strategy}\label{sec:analysis}

To fully exploit the fast timing of the LAPPD in selecting a Cherenkov-rich population of photons, the full time profile must be robustly modeled. This requires both precise measurement of the scintillation time profile of the sample under consideration, as well as an understanding of the ``trigger profile'', or the distribution of physical times between energy deposition in the sample and the triggering of the DAQ, which is itself influenced by the time profile of the sample. We describe below the importance of understanding the trigger profile, the formulation of an analytic fit used to measure the scintillation time profile, and the use of the MC to optimize a final Cherenkov selection cut.

\subsection{Trigger profile}

\begin{figure}[b!]
        \centering
        \includegraphics[width=0.45\textwidth]{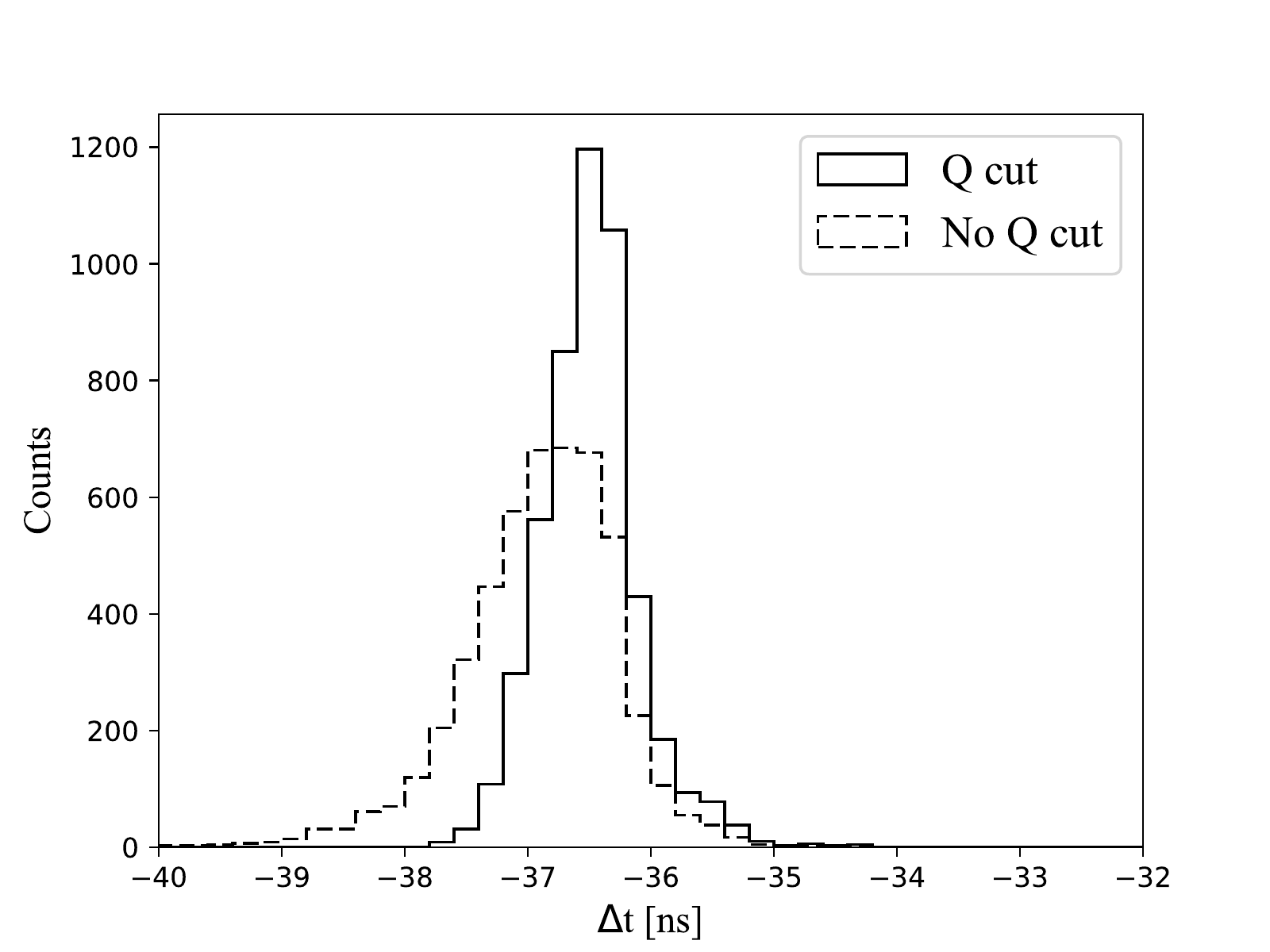}
        \caption{The observed time profile of Cherenkov light in the 1\% WbLS MC, before and after applying the trigger charge cut. The cut selects events that have high occupancy in the trigger PMT, which narrows and symmetrizes the trigger profile. The small tail at larger times is due to backscatters of PEs within the LAPPD.}
        \label{fig:trigger_q_cut}
    \end{figure}

Because the trigger PMT operates in a multi-PE regime, the threshold-crossing time described in Section \ref{sec:waveform_analysis}, associated with the first photon, is distributed according to the first order statistic of the time profile, with the associated sample size equal to the total number of detected photons. For low occupancies (one to a few photons), the first order statistic is asymmetric, which leads to a spilling of scintillation photons into the prompt region of the time profile, which reduces the Cherenkov purity in the prompt region. At higher occupancies, the trigger profile becomes symmetric and approximately Gaussian. This is also the higher energy regime, in which there is a more favorable Cherenkov-to-scintillation ratio. This effect is illustrated in Figure \ref{fig:trigger_q_cut}, which shows how the observed time profile of Cherenkov light in the MC changes due to the trigger charge cut.
    
The WbLS time profiles have previously been measured in the CHESS apparatus \cite{Caravaca:2020lfs} but, due in part to the MC-based fit method employed, and in part to shortcomings of the modeling of the trigger profile, did not achieve adequate sensitivity to the scintillator rise-time. Knowledge of the rise-time is critical to the modeling of the peak region where there is appreciable Cherenkov content. In this work, we first analytically fit for the scintillation time profile in the Gaussian trigger profile regime, as described below.

\subsection{Selection cuts}

After processing the raw waveforms and applying quality and muon-rejection criteria (see Section \ref{sec:waveform_analysis}), a charge cut is applied to the trigger PMT, which has two effects: the removal of Cherenkov-poor, low energy decays from the dataset, and the symmetrization of the trigger profile. The threshold charge values for the three samples considered in this work are listed in Table \ref{tab:charge_cuts}, along with the waveform-level cuts applied to each channel.

\begin{table}[t!]
    \small
    \centering
    \begin{tabular}{r c c c}
        \- &
            \textbf{1\%} &
            \textbf{5\%} &
           \textbf{10\%} \\
        \hline \hline 
        Trigger charge [pC] &
            $> 14.0$ &
            $> 18.1$ &
            $> 25.7$ \\
        Trigger amplitude [mV] &
            $> 15.0$ \\
        Signal amplitude [mV] &
            $> 5.0$ \\
        Muon panel amplitude [mV] &
            $< 10$ \\
        \# of threshold crossings &
            $< 2$ \\
    \end{tabular}
    \caption{Summary of selection cuts used to select SPE pulses in the LAPPD and timing PMT, and to ensure the trigger PMT operates with an approximately Gaussian profile. Only the trigger charge cut is sample-dependent.}
    \label{tab:charge_cuts}
\end{table}

\subsection{Analytic model}

The emission time profile of scintillation light from LAB-based liquid scintillators under $\beta^{-}$ excitation has been studied extensively, e.g. \cite{Caravaca:2020lfs,Anderson:2020xxb,OKeeffe:2011dex,Lombardi:2013nla,MarrodanUndagoitia:2009kq,Li2011}. In the present work, we assume a multiple exponential model modified to allow for a non-zero a rise-time, as defined in Birks \cite{birks}:

\begin{equation}\label{eq:scint_time_profile}
\begin{split}
    S(t)
        = \sum_{i=0}^{n}
            A_{i}\pp{
                \frac{e^{-t/\tau_{i}} - e^{-t/\tauR}}
                     {\tau_{i} - \tau_{R}}
                } \\
    \sum_{i=0}^{n} A_{i} = 1,
\end{split}
\end{equation}
where the $\tau_{i}$ and $A_{i}$ are the lifetimes and relative normalizations of the $n$ decay modes, and $\tauR$ is the rise-time of the scintillator. In this work, we follow \cite{Caravaca:2020lfs} and allow for $n = 2$ decay modes.

We model the production of Cherenkov light as instantaneous; an average electron deposits its full energy in 20~ps, as modeled in the MC, which is negligible when added in quadrature to the $O\pp{100~\text{ps}}$ resolution of the system. Because the trigger profile is effectively Gaussian, we model the distribution by either the LAPPD or timing PMT as:
\begin{equation}\label{eq:analytic_model}
\begin{split}
    F\pp{t}
        = \pp{1 - \fDarkj}G\pp{t - \tzj; \sigmaj} &\otimes \\ 
          \pp{
                \fCj\delta\pp{t}
              + \pp{1 - \fCj}S\pp{t}
            }
          &+ \frac{\fDarkj}{T}, \\
\end{split}
\end{equation}
where $j$ specifies the device (either LAPPD or timing PMT) and $G$ denotes a Gaussian, $\sigma$ is the effective resolution of the system, $\tz$ is the overall system delay, $\fDark$ is the fraction of the data which is comprised of dark hits, $\fC$ is the fraction of detected photons that are Cherenkov, and $T$ is the size of the analysis window, which defines the probability density of uniformly distributed dark hits, and is 100~ns in this work. The model describes the Cherenkov and scintillation time profiles convolved with a Gaussian system response; $T$ is a fixed parameter, with all other parameters free. A joint model, describing both the LAPPD and timing PMT datasets, is defined by applying the model to the two respective devices, with the four parameters defining the scintillation time profile common to the two.

After applying selection cuts, the data for both the LAPPD and timing PMT are binned into histograms each with bin width of 100~ps. The joint model is fit to the dataset by minimizing the negative joint binned log-likelihood, and uncertainties on all parameters are computed by profiling the likelihood function.

\subsection{Monte Carlo comparison and Cherenkov selection}

The scintillation time profiles found from the analytic fits described above are used as input to the MC simulation, described in Section \ref{sec:simulation}. This provides a verification of the results of the fits, and allows for improvements in modeling due to non-Gaussian effects arising from residual asymmetry of the trigger profile and backscatters of photoelectrons in the LAPPD. The MC also includes additional optical effects, such as absorption and reemission of Cherenkov and scintillation photons, which constitute futher improvements to the model.

The total numbers of Cherenkov and scintillation photons produced in the MC is determined by the index of refraction and scintillation light yield, respectively, which are both inputs to the simulation. This is in contrast to the floating normalization allowed in the analytic fit (via $\fC$). To evaluate the MC with the best-fit time profile, a scale factor applied to the scintillation light yield is iterated over, changing the light output in steps of 15 photons/MeV, and the value which best models the observed timing distribution is used as an effective parameter of the MC model. As the goal is to demonstrate the achievable Cherenkov-purity inherent to WbLS, dark hits are not simulated and are statistically subtracted from the histograms observed in data when comparing to the MC; this is achieved by fitting a flat line to the region before the prompt light. The binned comparison is performed across a 140~ns window, with a bin width of 100~ps. Bins with zero entries in either the data or the MC are not included in the $\chi^{2}$ calculation.

For any MC evaluation, the Cherenkov purity of a prompt time window is defined as:
\begin{equation}\label{eq:purity}
    P
        = \frac{\int_{-\infty}^{\tf} C\pp{t} \differential{t}}
               {\int_{-\infty}^{\tf} C\pp{t} + S\pp{t} \differential{t}},
\end{equation}
where $C$ and $S$ are the populations of Cherenkov and scintillation light, respectively, and $t_{f}$ is the upper edge of the time window. To maximize the purity of Cherenkov photons while retaining significant statistics, we define:
\begin{equation}\label{eq:r}
    R\pp{\tf}
        = P\pp{\tf} \times \int_{-\infty}^{\tf} C\pp{t} \differential{t},
\end{equation}
and determine an optimal $\tf$ by maximizing $R$. This is equivalent to optimizing the standard signal-to-background metric $C/\sqrt{C + S}$, but through a more intuitive quantity, as $R$ is explicitly constructed from the purity and total number of Cherenkov photons.

\section{Results}\label{sec:results}

\subsection{Analytic fits}\label{sec:fit_results}

\begin{figure*}[t!]
    \centering
    \includegraphics[width=0.325\textwidth]{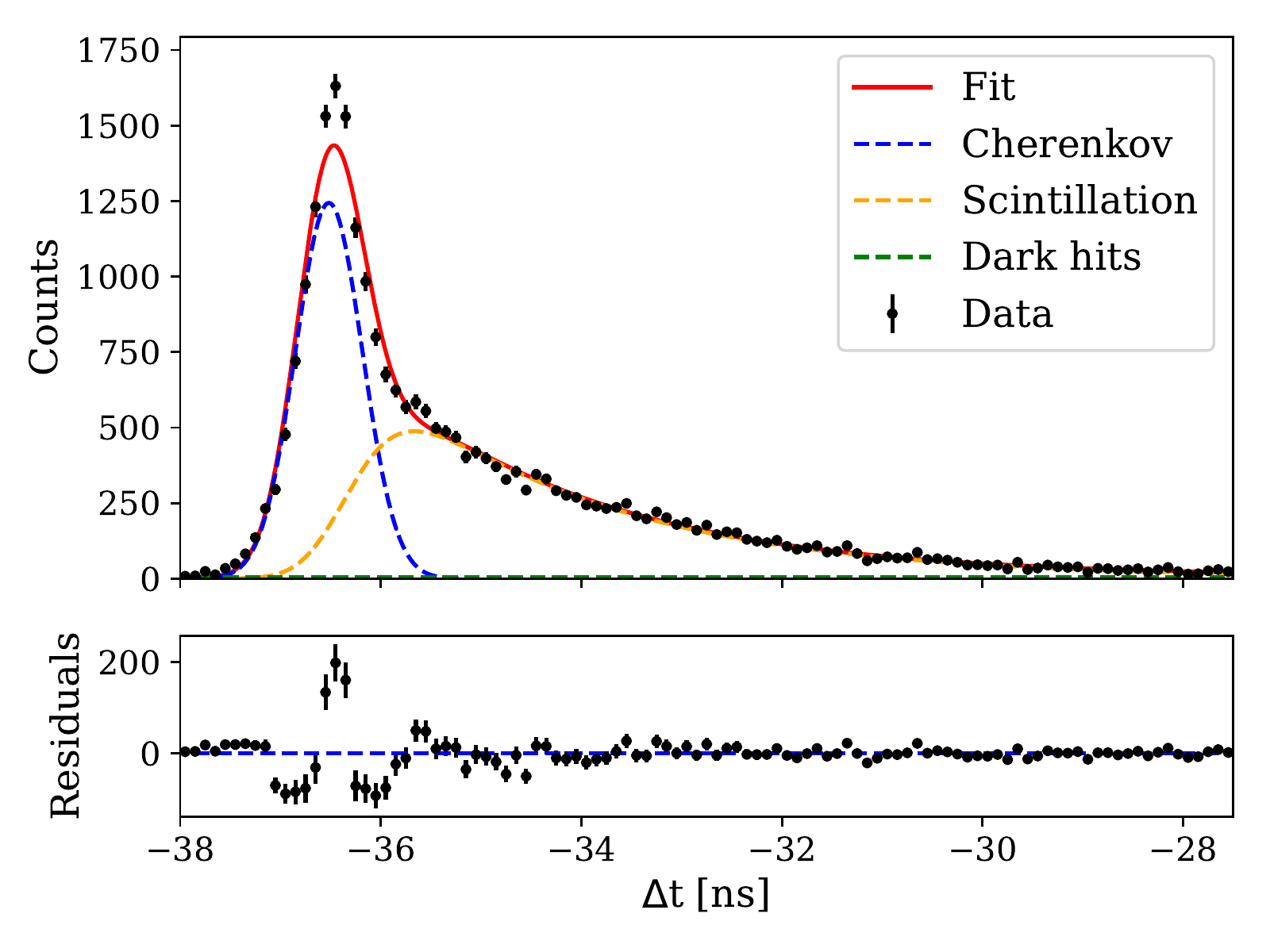}
    \includegraphics[width=0.325\textwidth]{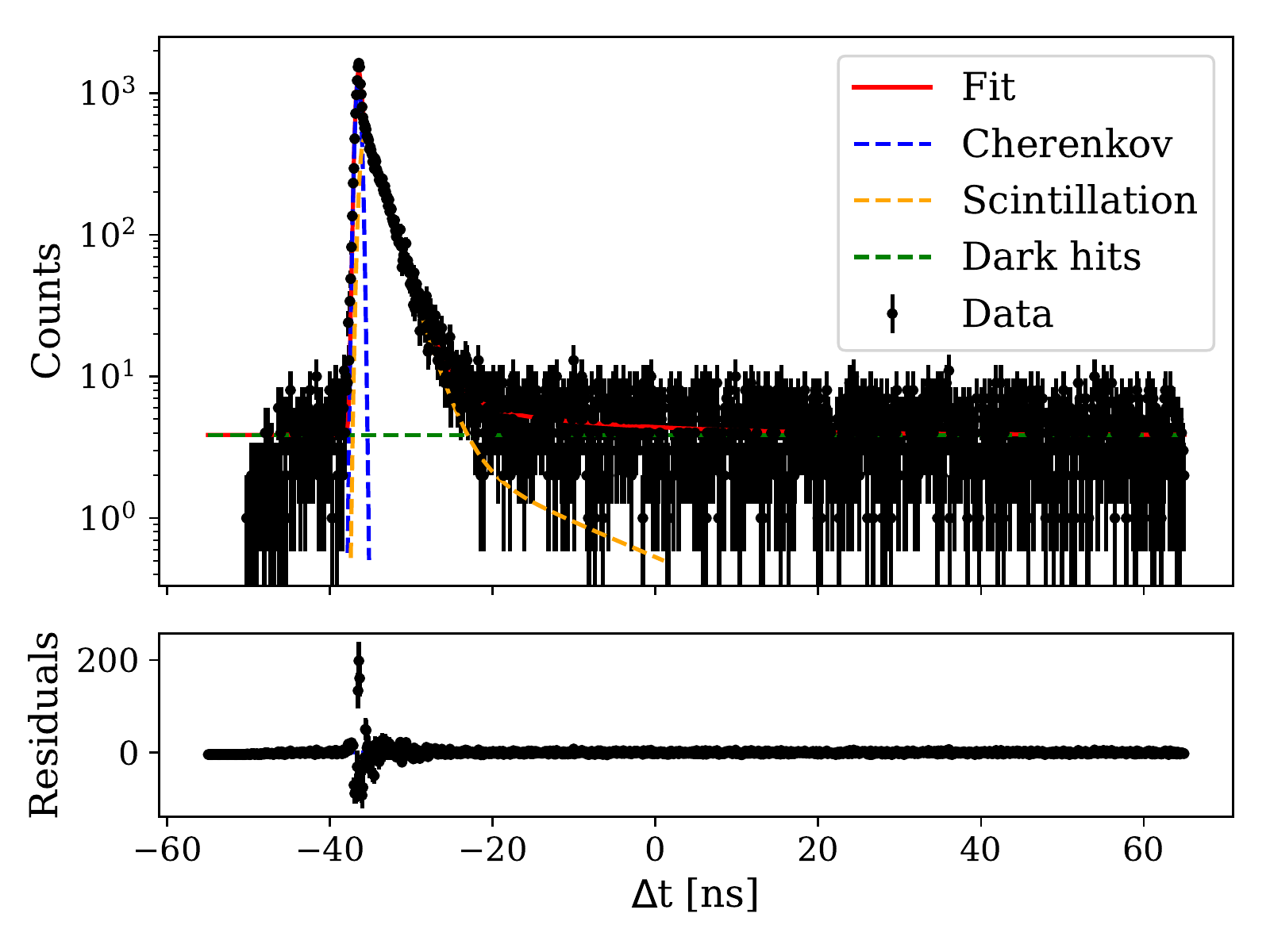}
    \includegraphics[width=0.325\textwidth]{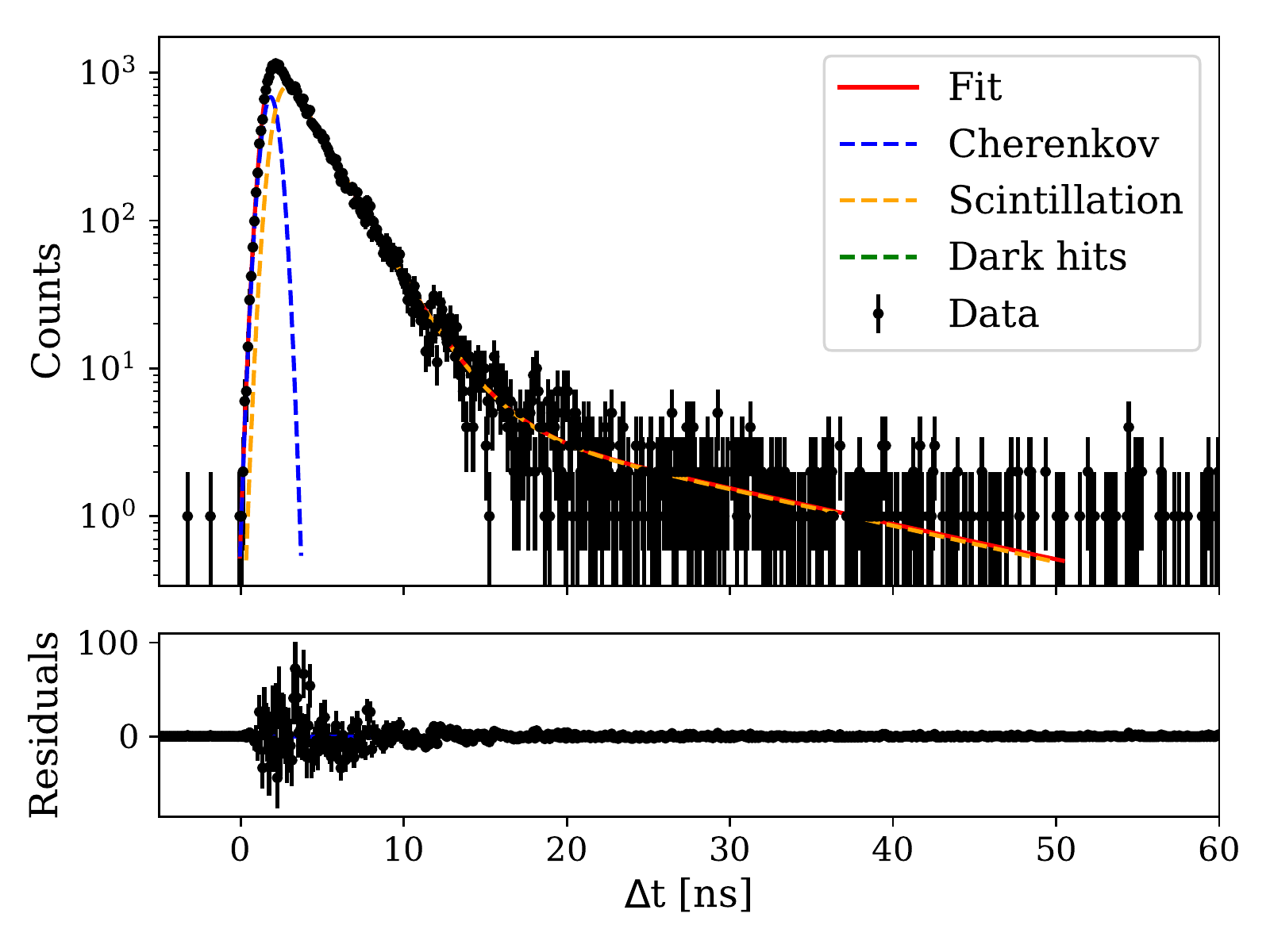}
    \caption{The best-fit analytic model compared to the 1\% WbLS data, for the peak-region of the LAPPD (left), full analysis window of the LAPPD (middle), and full analysis window of the timing PMT (right). The analytical model contains several approximations and is improved upon using our full MC (Figure \ref{fig:mc_comparison}, left).}
    \label{fig:wbls1pct_fit}
\end{figure*}

\begin{figure*}[t!]
    \centering
    \includegraphics[width=0.325\textwidth]{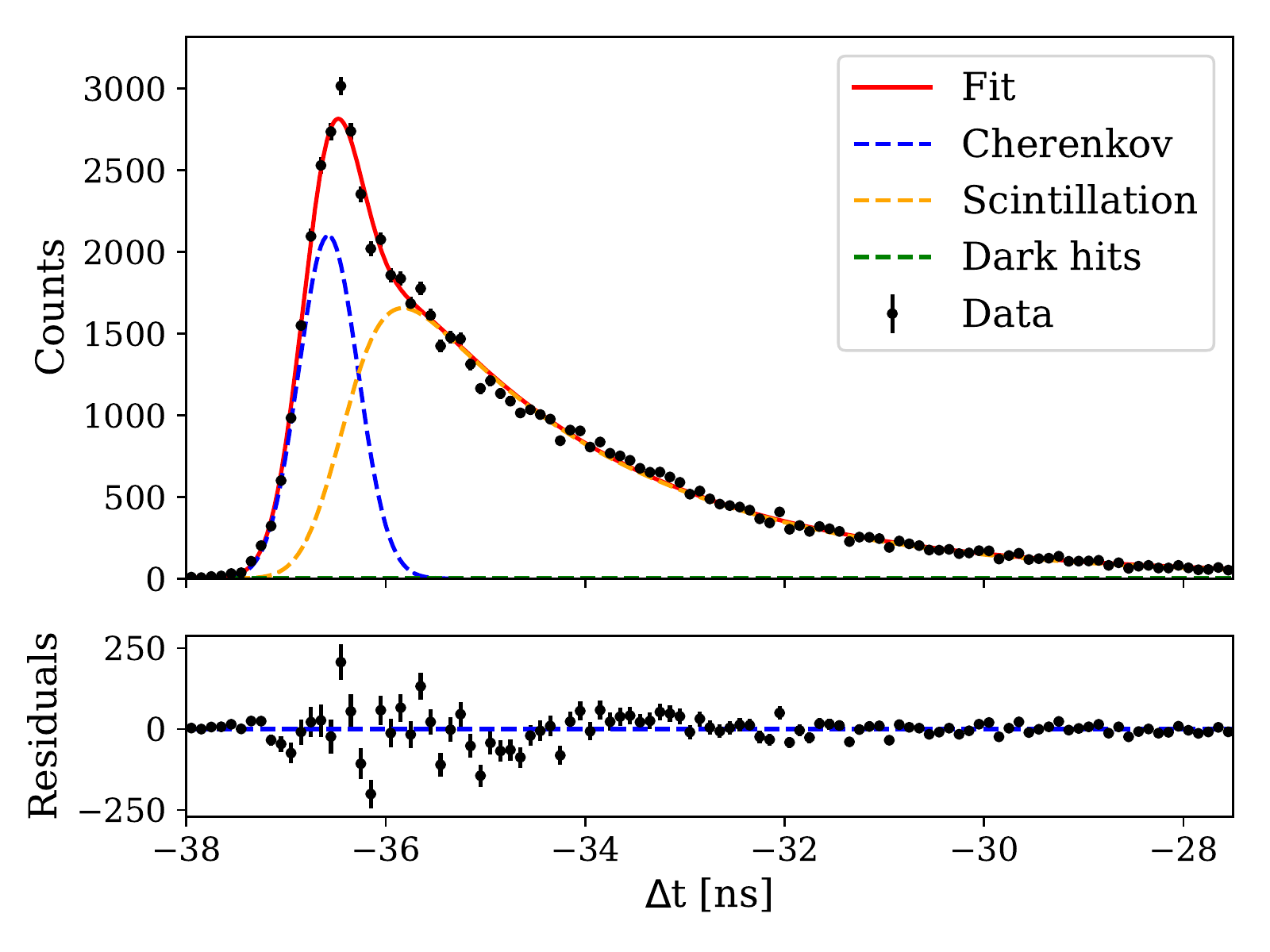}
    \includegraphics[width=0.325\textwidth]{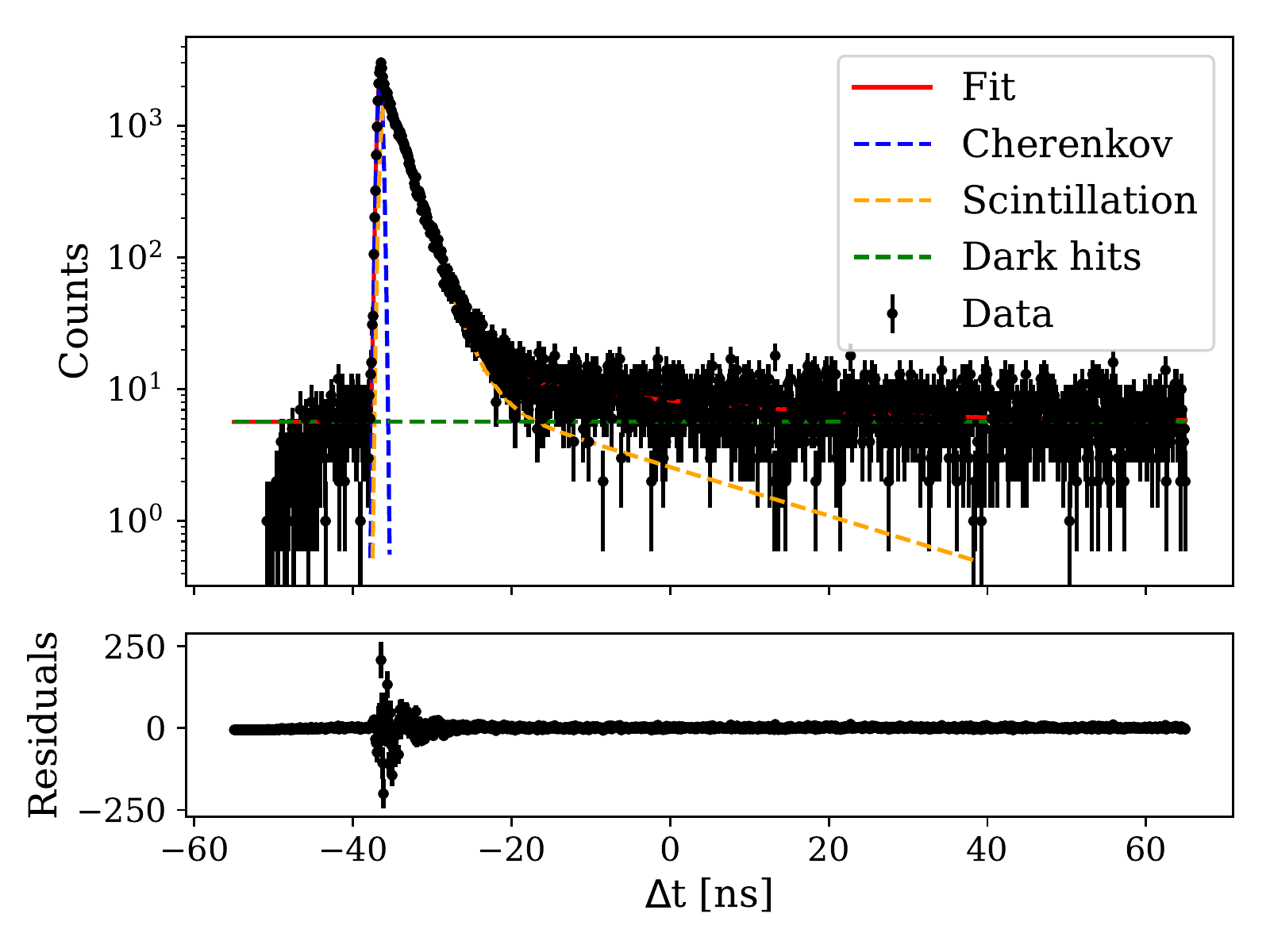}
    \includegraphics[width=0.325\textwidth]{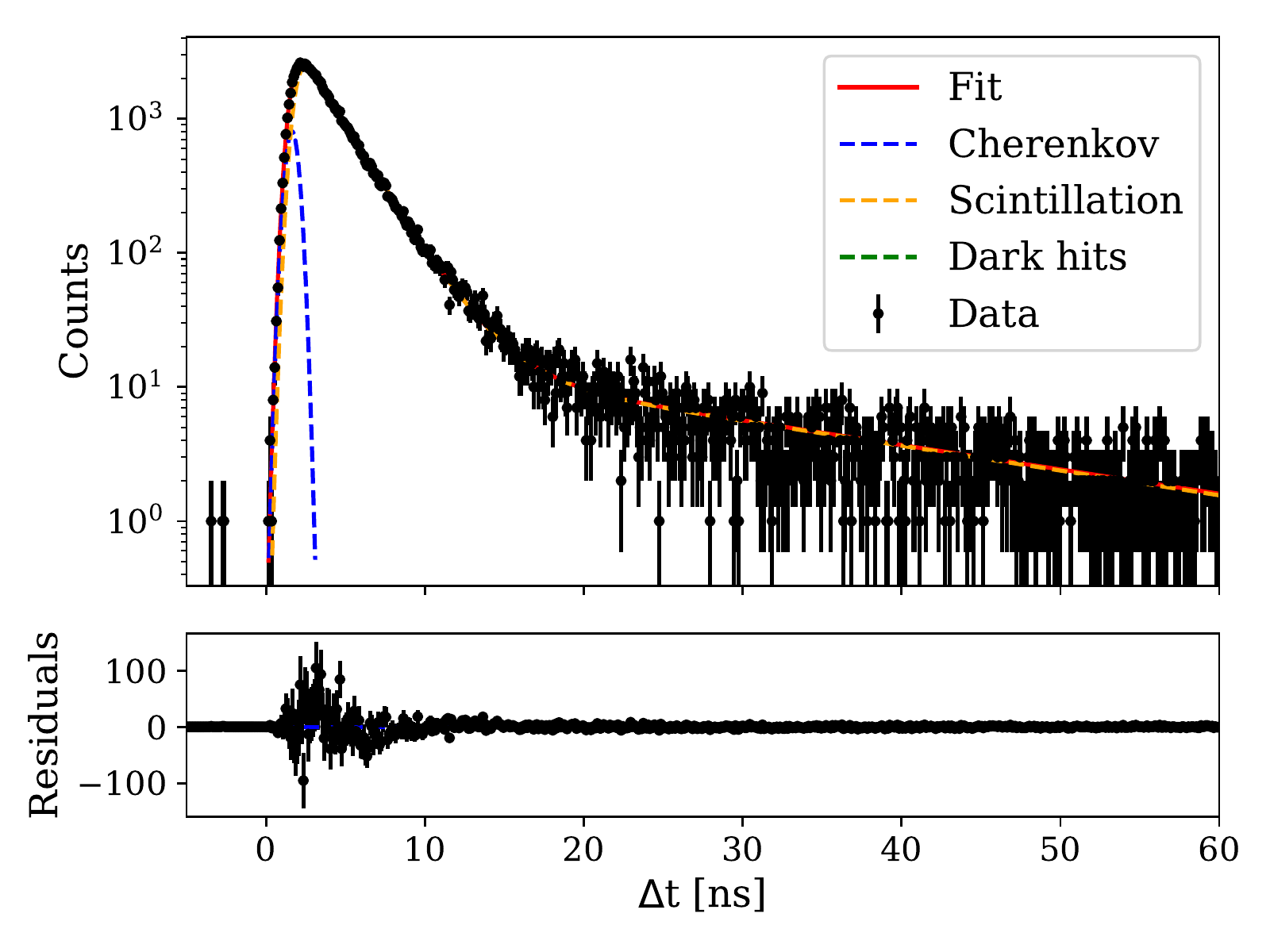}
    \caption{The best-fit analytic model compared to the 5\% WbLS data, for the peak-region of the LAPPD (left), full analysis window of the LAPPD (middle), and full analysis window of the timing PMT (right). The analytical model contains several approximations and is improved upon using our full MC (Figure \ref{fig:mc_comparison}, center).}
    \label{fig:wbls5pct_fit}
\end{figure*}

\begin{figure*}[t!]
    \centering
    \includegraphics[width=0.325\textwidth]{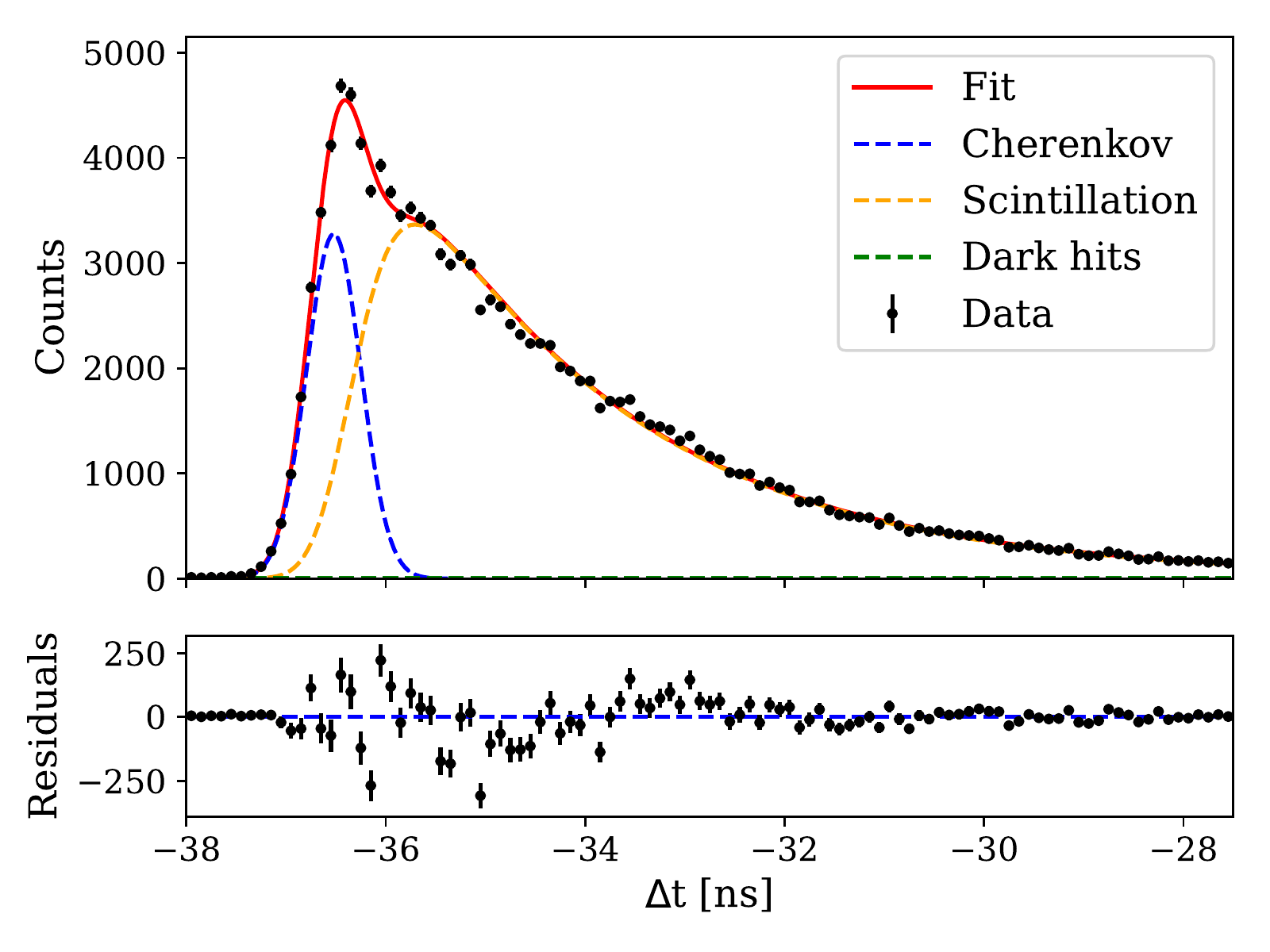}
    \includegraphics[width=0.325\textwidth]{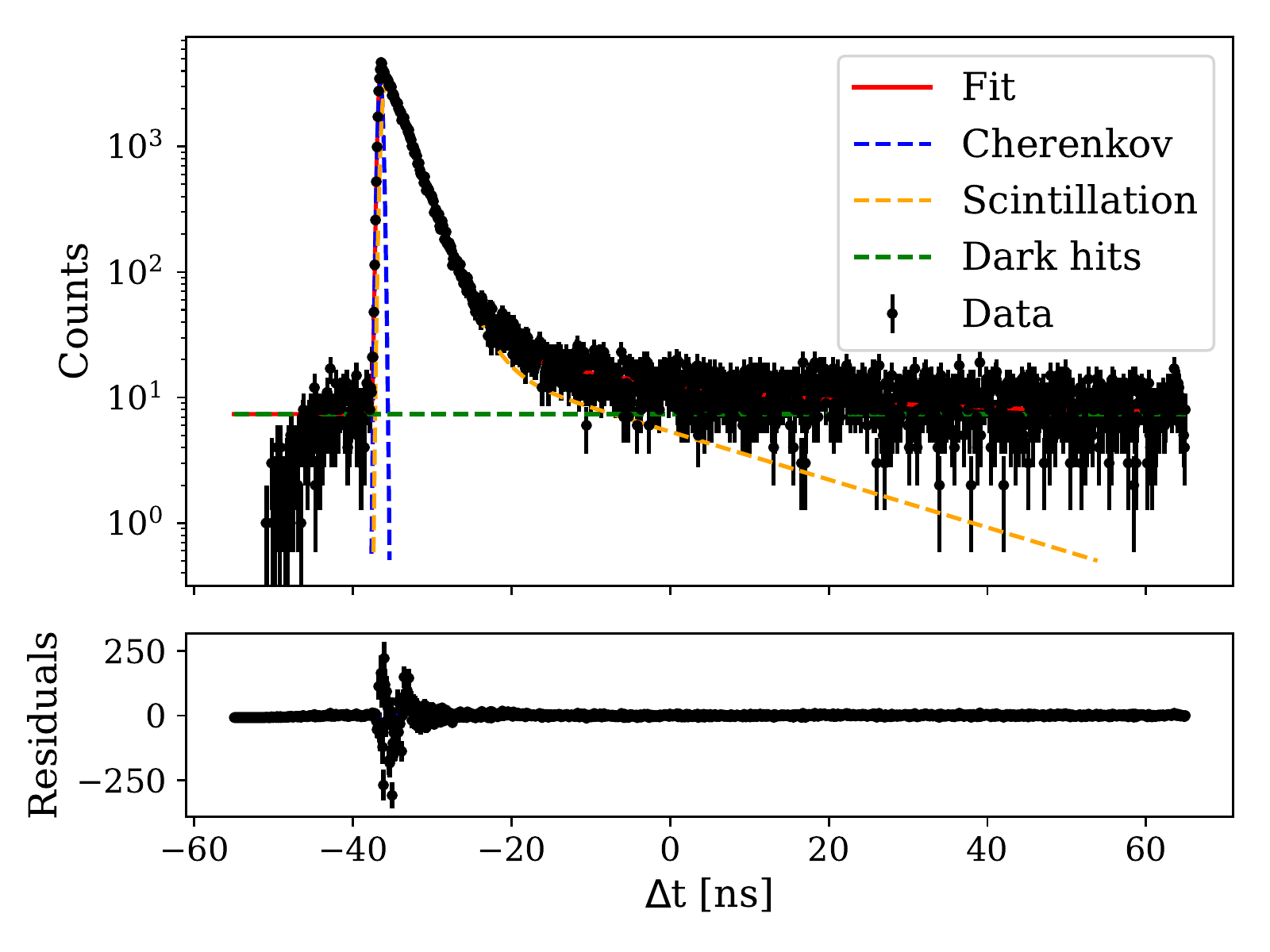}
    \includegraphics[width=0.325\textwidth]{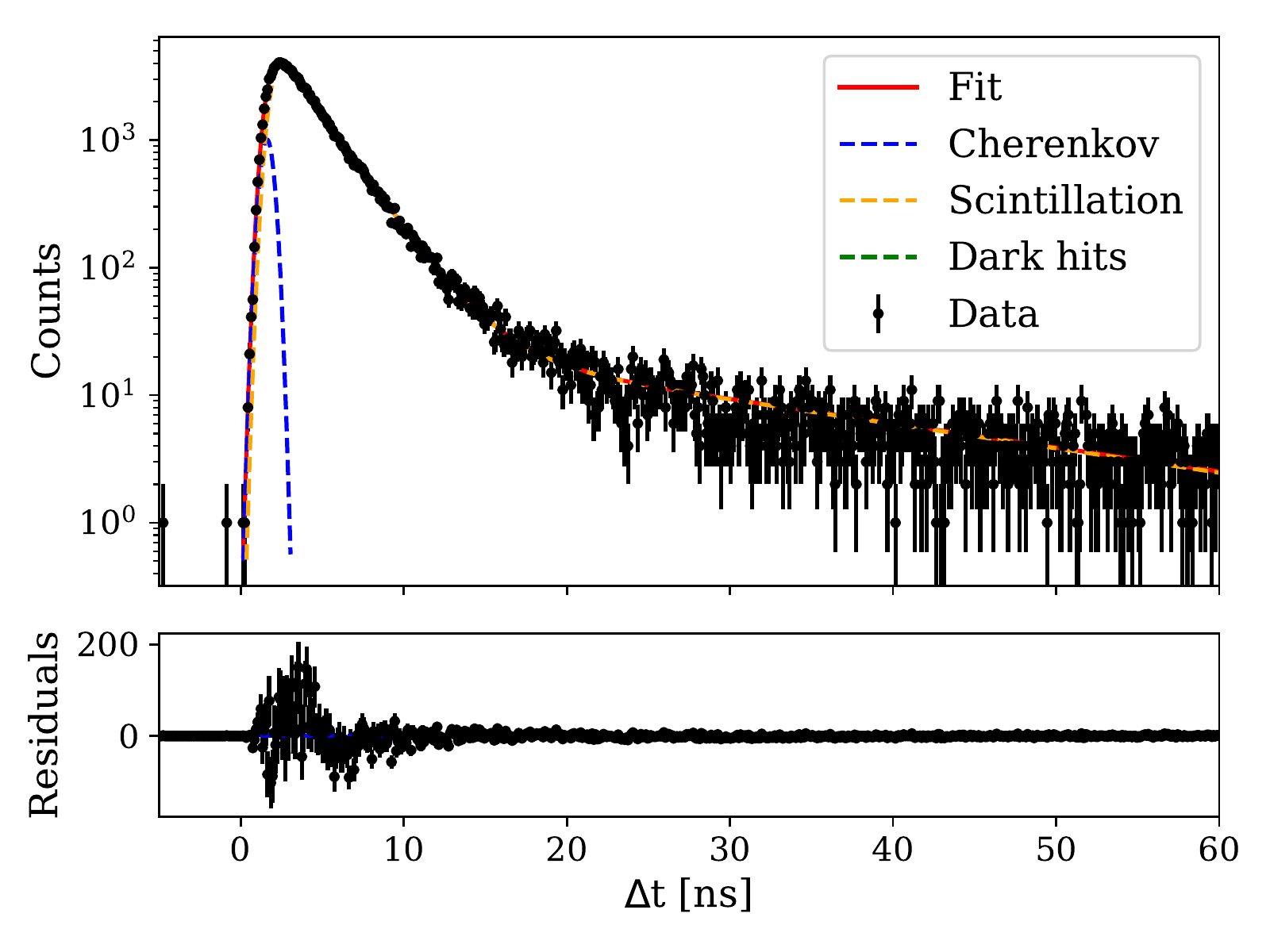}
    \caption{The best-fit analytic model compared to the 10\% WbLS data, for the peak-region of the LAPPD (left), full analysis window of the LAPPD (middle), and full analysis window of the timing PMT (right). The analytical model contains several approximations and is improved upon using our full MC (Figure \ref{fig:mc_comparison}, right).}
    \label{fig:wbls10pct_fit}
\end{figure*}

The results of the analytic fits are shown in Table \ref{tab:fit_results} and Figures \ref{fig:wbls1pct_fit}, \ref{fig:wbls5pct_fit}, and \ref{fig:wbls10pct_fit}, and are generally consistent with the previous MC-based measurement \cite{Caravaca:2020lfs}. The model generally predicts the features apparent in the data, namely the rising edge of prompt Cherenkov light, transition to the scintillation-dominant regime, and scintillation tail, but underpredicts the peak region at low scintillation concentrations. This is likely due to the unmodeled interplay between residual asymmetry in the trigger profile and late pulsing in the LAPPD. Improvements to the modeling of this region are achieved with the full simulation (Section \ref{sec:mc_results}). The effective resolutions determined by the fit systematically decrease with higher scintillator concentrations, due to the higher occupancies of the trigger PMT under relatively similar scintillation time profiles.

\begin{table}[t!]
    \small
    \centering
    \begin{tabular}{r c c c}
        \- &
            \textbf{1\%} &
            \textbf{5\%} &
           \textbf{10\%} \\
        \hline
        \hline
        $\tau_{R}~\text{[ps]}$      &
            $270\plusminus{26}{20}$ &
            $209\plusminus{10}{11}$ &
            $276\plusminus{7}{7}$ \\
        $\tau_{1}~\text{[ns]}$      &
            $2.22\plusminus{0.02}{0.02}$ &
            $2.25\plusminus{0.01}{0.01}$ &
            $2.36\plusminus{0.01}{0.01}$ \\
        $\tau_{2}~\text{[ns]}$      &
            $17.7\plusminus{1.3}{1.1}$ &
            $23.5\plusminus{1.0}{0.9}$ &
            $22.8\plusminus{0.7}{0.7}$ \\
        $A_{1}~\text{[\%]}$         &
            $95.6\plusminus{0.3}{0.3}$ &
            $94.8\plusminus{0.1}{0.1}$ &
            $94.9\plusminus{0.1}{0.1}$ \\
        $\sLAPPD~\text{[ps]}$       &
            $334\plusminus{4}{4}$ &
            $298\plusminus{4}{3}$ &
            $273\plusminus{3}{3}$ \\
        $\sPMT~\text{[ps]}$         &
            $495\plusminus{13}{7}$ &
            $381\plusminus{8}{5}$ &
            $372\plusminus{5}{4}$ \\
        \hline
        $\chi^{2}/\ndf$ &
            $2968/2388$ &
            $3031/2388$ &
            $3373/2388$ \\
    \end{tabular}
    \caption{The fit results for the WbLS mixtures. The scintillation emission timing parameters are defined in Equation \ref{eq:analytic_model}.}
    \label{tab:fit_results}
\end{table}

\subsection{MC Comparison and Cherenkov Selection}\label{sec:mc_results}

\begin{figure*}[t!]
    \centering
    \includegraphics[width=0.325\textwidth]{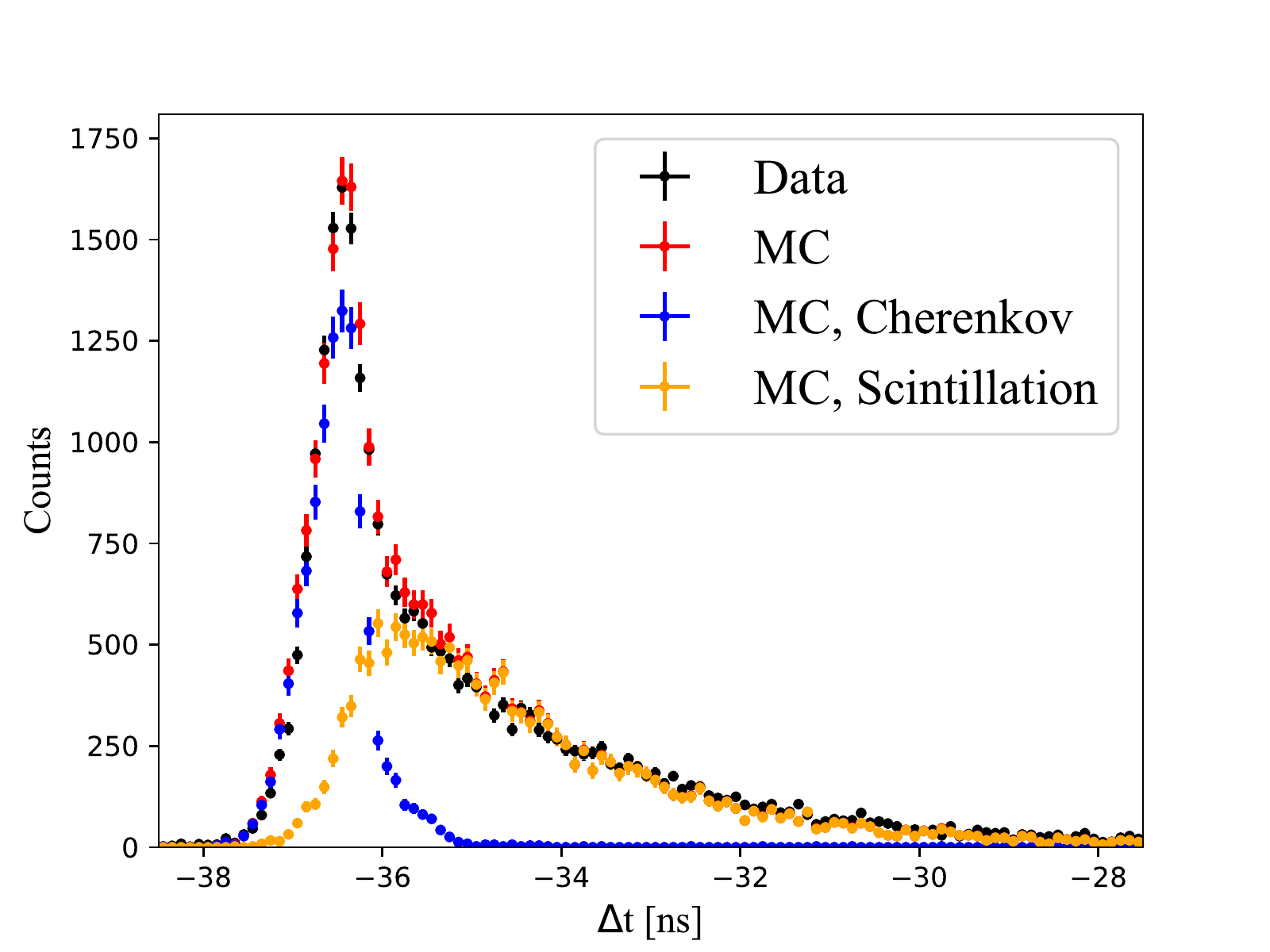}
    \includegraphics[width=0.325\textwidth]{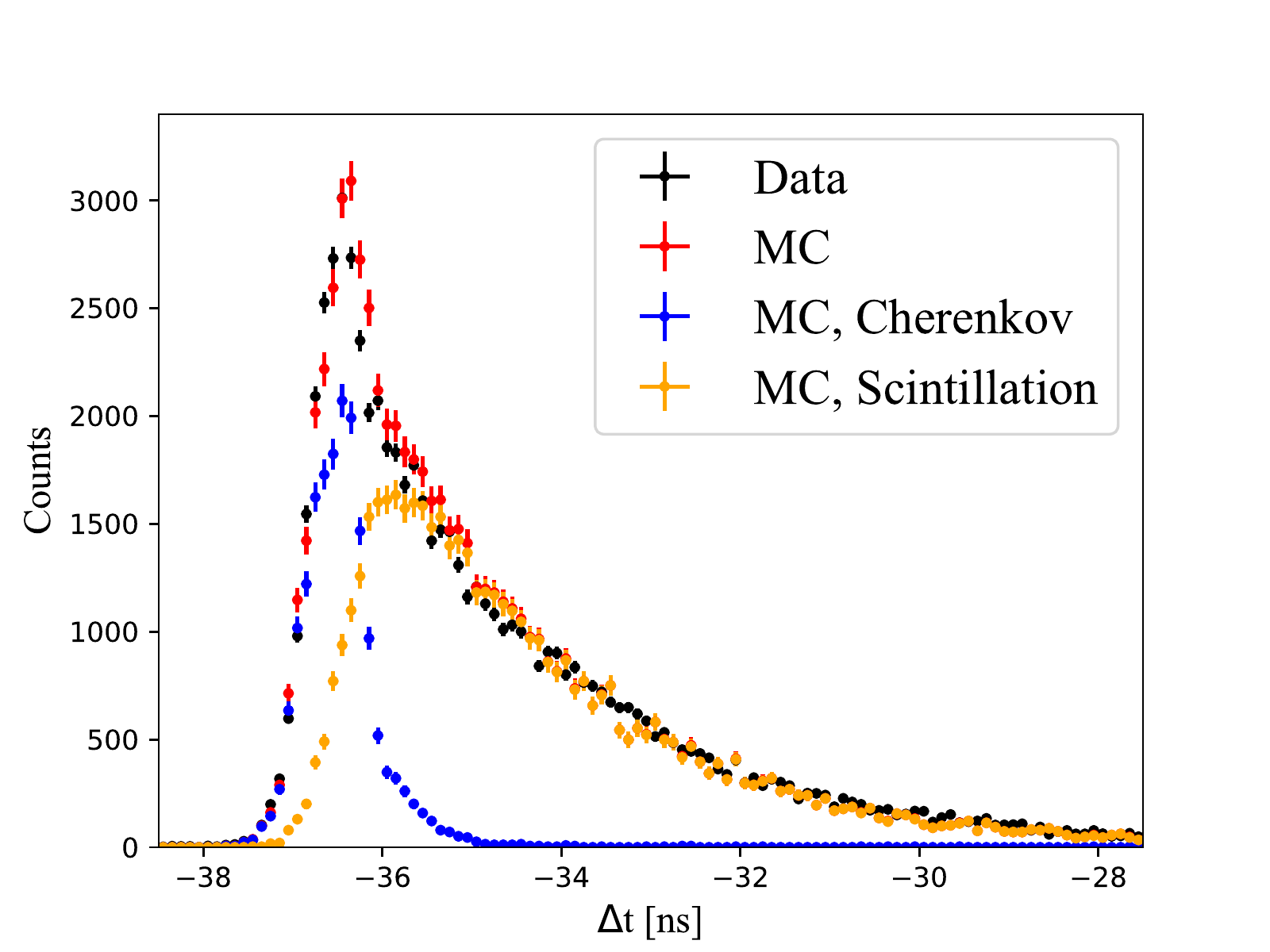}
    \includegraphics[width=0.325\textwidth]{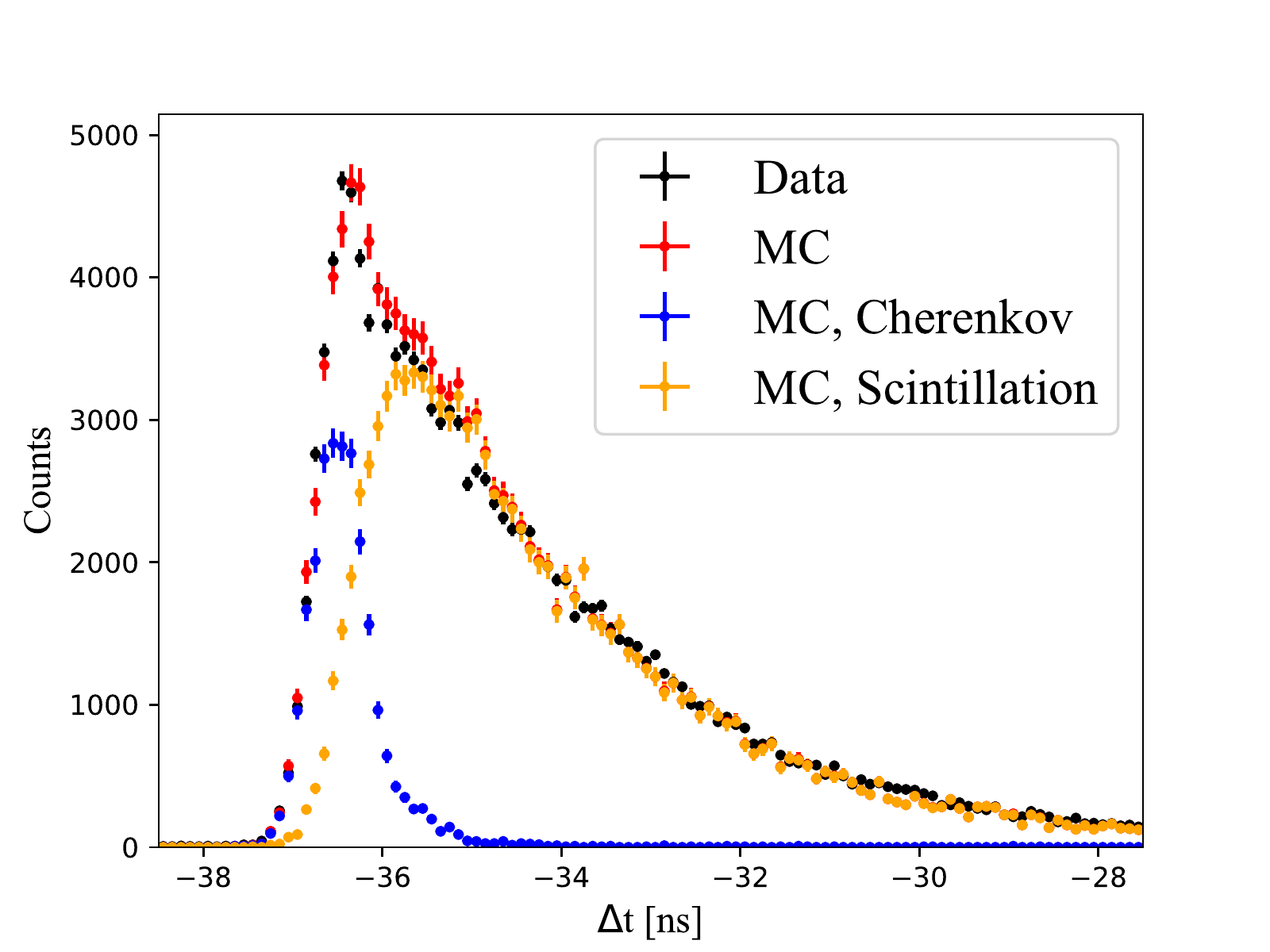}
    \caption{The timing of the best-fit simulation model compared to that observed in data for the 1\% (left), 5\% (middle), and 10\% (right) WbLS samples. Dark hits were not simulated and have been statistically subtracted out of the data. The MC model shows improved agreement with the data, relative to the analytic fits, in the Cherenkov peak.}
    \label{fig:mc_comparison}
\end{figure*}

The timing distributions generated by MC using the best-fit scintillation time profile and optimal light yield are compared to the data in Figure \ref{fig:mc_comparison}. The optimal prompt time cuts and resulting Cherenkov purities are reported in Table \ref{tab:mc_results}. The simulation improves on the analytic model in the peak region by including non-Gaussian effects due to residual asymmetry in the trigger profile and late pulsing in the LAPPD, as well subtle distortions due to the optics of the setup. In all cases, the optimal Cherenkov selection window extends to is 200 or 300~ps after to the peak of the timing distribution, with the purity of Cherenkov photons in this window being greater than 60\%. The 1\% sample, with a scintillation light yield on the order of 200~photons/MeV, admits a purity of greater than 80\%.

\begin{table}[b!]
    \small
    \centering
    \begin{tabular}{r c c c}
        \- &
           \textbf{1\%} &
           \textbf{5\%} &
          \textbf{10\%} \\
          \hline
          \hline
        $\tf~\text{[ps]}$ &
            300 &
            200 &
            200 \\
        $P~\text{[\%]}$ &
            80.4 &
            68.6 &
            64.3 \\
        \hline
        $\chi^{2}/\ndf$ &
            298.1/235 &
            561.7/457 &
            698.8/505 \\
    \end{tabular}
    \caption{The Cherenkov selection results for the WbLS mixtures.}
    \label{tab:mc_results}
\end{table}

\section{Conclusion}

This work constitutes the first demonstration of the combination of WbLS and LAPPDs to detect Cherenkov light from a scintillating material with high purity. The experimental apparatus admits an effective system resolution $O\pp{100~\text{ps}}$. Using both an LAPPD and a conventional PMT, the scintillation time profiles of three WbLS samples, loaded with liquid scintillator at the 1\%, 5\%, and 10\% levels, were measured under an analytic model, with improved sensitivity to the rise-time, and are in general agreement with previous measurements of the decay modes. Using an MC model of the experimental geometry and electronics, the results of the analytic model were verified, and optimal Cherenkov-selection windows were developed exclusively on the basis of timing. The resulting purities of Cherenkov light are systematically larger than 60\%, with 1\% WbLS exhibiting greater than 80\% purity. 

Previous measurements of the same WbLS mixtures, in the same low energy regime, utilized an array of 1-inch PMTs to characterize the time profiles of the WbLS samples \cite{Caravaca:2020lfs}. In that work, direct separation of the different photon populations in the low energy regime was not attempted, but instead demonstrated using high energy muons. In contrast, this work demonstrates the effectiveness of fast photodetectors, such as the LAPPD, in achieving high Cherenkov purity at the MeV scale. This capability is advantangeous to achieve the physics goals of a hybrid detector such as \theia{} \cite{Askins:2019oqj}, several of which depend on direction reconstruction using Cherenkov light.

Additionally, these measurements verify the scintillation model used in large-scale simulations, reinforcing the studies in \cite{Land:2020oiz}, where the impact of WbLS and LAPPDs on reconstruction in multi-ktonne detectors is evaluated. For example, it's shown that detectors which utilize LAPPDs achieve enhanced angular resolution compare to those equipped with conventional PMTs. Specifically, the natural degradation in angular resolution with increasing scintillator concentration is largely mitigated by the fast timing of the LAPPDs, which allows for improved energy and vertex reconstruction while retaining meaningful directional information. This is demonstrated to translate to suppression of \isotope{B}{8} solar neutrinos background for $0\nu\beta\beta$ searches in a pure LS detector, leading to increased sensitivity. For a 50 ktonne WbLS detector equipped with LAPPDs, it's shown that the CNO neutrino flux could be measured to 10\% uncertainty after five years of data-taking, which would improve on the recent measurement made by Borexino \cite{BOREXINO:2020aww}.

This work demonstrates the achievement of significant Cherenkov purity in WbLS samples measured using LAPPDs, which advances the development of upcoming optical neutrino detectors, including ANNIE, AIT-NEO, and \theia{}. Future efforts, using the same LAPPD-based apparatus, to investigate PID using WbLS, and measure the timing of various slow scintillators and loaded WbLS mixtures are anticipated.

\section{Acknowledgements}

The authors thank Benjamin Land and Javier Caravaca for their early contributions in the development of the LAPPD setup, \texttt{RAT-PAC} simulation, and WbLS optical model, which are critical for this work, and thank Incom technology for providing the LAPPD device, and for support and advice on its operation.

This material is based upon work supported by the U.S. Department of Energy, Office of Science, Office of High Energy Physics, under Award Number  DE-SC0018974.  Work conducted at Lawrence Berkeley National Laboratory was performed under the auspices of the U.S. Department of Energy under Contract DE-AC02-05CH11231. The work conducted at Brookhaven National Laboratory was supported by the U.S. Department of Energy under contract DE-AC02-98CH10886. EJC was funded by the Consortium for Monitoring, Technology, and Verification under Department of Energy National Nuclear Security Administration award number DE-NA0003920. The project was funded by the U.S. Department of Energy, National Nuclear Security Administration, Office of Defense Nuclear Nonproliferation  Research and Development (DNN R\&D).

\clearpage

\bibliographystyle{ieeetr}
\bibliography{main.bib}

\begin{thebibliography}{10}

\bibitem{IMB}
R.~Becker-Szendy {\em et~al.}, ``{Neutrino measurements with the IMB
  detector},'' {\em Nucl. Phys. Proc. Suppl.}, vol.~38, pp.~331--336, 1995.

\bibitem{Ahn:2006zza}
M.~H. Ahn {\em et~al.}, ``{Measurement of Neutrino Oscillation by the K2K
  Experiment},'' {\em Phys. Rev.}, vol.~D74, p.~072003, 2006.

\bibitem{Ahmad:2002jz}
Q.~R. Ahmad {\em et~al.}, ``{Direct evidence for neutrino flavor transformation
  from neutral current interactions in the Sudbury Neutrino Observatory},''
  {\em Phys. Rev. Lett.}, vol.~89, p.~011301, 2002.

\bibitem{Fukuda:1998mi}
Y.~Fukuda {\em et~al.}, ``{Evidence for oscillation of atmospheric
  neutrinos},'' {\em Phys. Rev. Lett.}, vol.~81, pp.~1562--1567, 1998.

\bibitem{kamland}
K.~Eguchi {\em et~al.}, ``{First results from KamLAND: Evidence for reactor
  anti-neutrino disappearance},'' {\em Phys. Rev. Lett.}, vol.~90, p.~021802,
  2003.

\bibitem{Seo:2019shs}
H.~Seo, ``{Recent Result from RENO},'' {\em J. Phys. Conf. Ser.}, vol.~1216,
  no.~1, p.~012003, 2019.

\bibitem{An:2016ses}
F.~P. An {\em et~al.}, ``{Measurement of electron antineutrino oscillation
  based on 1230 days of operation of the Daya Bay experiment},'' {\em Phys.
  Rev. D}, vol.~95, no.~7, p.~072006, 2017.

\bibitem{Abe:2019vii}
K.~Abe {\em et~al.}, ``{Constraint on the matter\textendash{}antimatter
  symmetry-violating phase in neutrino oscillations},'' {\em Nature}, vol.~580,
  no.~7803, pp.~339--344, 2020.
\newblock [Erratum: Nature 583, E16 (2020)].

\bibitem{Adamson:2017gxd}
P.~Adamson {\em et~al.}, ``{Constraints on Oscillation Parameters from $\nu_e$
  Appearance and $\nu_\mu$ Disappearance in NOvA},'' {\em Phys. Rev. Lett.},
  vol.~118, no.~23, p.~231801, 2017.

\bibitem{IceCube:2018cha}
M.~G. Aartsen {\em et~al.}, ``{Neutrino emission from the direction of the
  blazar TXS 0506+056 prior to the IceCube-170922A alert},'' {\em Science},
  vol.~361, no.~6398, pp.~147--151, 2018.

\bibitem{Aguilar-Arevalo:2018gpe}
A.~A. Aguilar-Arevalo {\em et~al.}, ``{Significant Excess of ElectronLike
  Events in the MiniBooNE Short-Baseline Neutrino Experiment},'' {\em Phys.
  Rev. Lett.}, vol.~121, no.~22, p.~221801, 2018.

\bibitem{Agostini:2020mfq}
M.~Agostini {\em et~al.}, ``{Experimental evidence of neutrinos produced in the
  CNO fusion cycle in the Sun},'' {\em Nature}, vol.~587, pp.~577--582, 2020.

\bibitem{Andringa:2015tza}
S.~Andringa {\em et~al.}, ``{Current Status and Future Prospects of the SNO+
  Experiment},'' {\em Adv. High Energy Phys.}, vol.~2016, p.~6194250, 2016.

\bibitem{Agostini:2021bxc}
M.~Agostini {\em et~al.}, ``{First demonstration of directional measurement of
  sub-MeV solar neutrinos in a liquid scintillator detector with Borexino},'' 9
  2021.

\bibitem{snoplus_angular}
{E. Fletcher}, ``{Angular Resolution and Directionality in SNO+},'' Master's
  thesis, {Queen's University}, 2019.

\bibitem{Jinping:2016iiq}
J.~F. Beacom {\em et~al.}, ``{Physics prospects of the Jinping neutrino
  experiment},'' {\em Chin. Phys. C}, vol.~41, no.~2, p.~023002, 2017.

\bibitem{YehWbLS}
M.~Yeh, S.~Hans, W.~Beriguete, R.~Rosero, L.~Hu, R.~Hahn, M.~Diwan, D.~Jaffe,
  S.~Kettell, and L.~Littenberg, ``A new water-based liquid scintillator and
  potential applications,'' {\em Nuclear Instruments and Methods in Physics
  Research Section A-accelerators Spectrometers Detectors and Associated
  Equipment - NUCL INSTRUM METH PHYS RES A}, vol.~660, pp.~51--56, 12 2011.

\bibitem{Bonventre:2018hyd}
R.~Bonventre and G.~D. Orebi~Gann, ``{Sensitivity of a low threshold
  directional detector to CNO-cycle solar neutrinos},'' {\em Eur. Phys. J. C},
  vol.~78, no.~6, p.~435, 2018.

\bibitem{Elagin:2016zgp}
A.~Elagin, H.~Frisch, B.~Naranjo, J.~Ouellet, L.~Winslow, and T.~Wongjirad,
  ``{Separating Double-Beta Decay Events from Solar Neutrino Interactions in a
  Kiloton-Scale Liquid Scintillator Detector By Fast Timing},'' {\em Nucl.
  Instrum. Meth.}, vol.~A849, pp.~102--111, 2017.

\bibitem{Aberle:2013jba}
C.~Aberle, A.~Elagin, H.~J. Frisch, M.~Wetstein, and L.~Winslow, ``{Measuring
  Directionality in Double-Beta Decay and Neutrino Interactions with
  Kiloton-Scale Scintillation Detectors},'' {\em JINST}, vol.~9, p.~P06012,
  2014.

\bibitem{Biller:2013wua}
S.~D. Biller, ``{Probing Majorana neutrinos in the regime of the normal mass
  hierarchy},'' {\em Phys. Rev.}, vol.~D87, no.~7, p.~071301, 2013.

\bibitem{WbLStalk}
M.~Yeh, ``{Water-based Liquid Scintillator}.''
  \url{https://indico.fnal.gov/event/3356/contributions/81175/attachments/51328/61443/Water-based_Liquid_Scintillator_Detector_RD_workshop_M._Yeh.pdf},
  2010.

\bibitem{Beacom:2003nk}
J.~F. Beacom and M.~R. Vagins, ``{GADZOOKS! Anti-neutrino spectroscopy with
  large water Cherenkov detectors},'' {\em Phys. Rev. Lett.}, vol.~93,
  p.~171101, 2004.

\bibitem{Kaptanoglu:2018sus}
T.~Kaptanoglu, M.~Luo, and J.~Klein, ``{Cherenkov and Scintillation Light
  Separation Using Wavelength in LAB Based Liquid Scintillator},'' {\em JINST},
  vol.~14, no.~05, p.~T05001, 2019.

\bibitem{Kaptanoglu:2017jxo}
T.~Kaptanoglu, ``{Characterization of the Hamamatsu 8" R5912-MOD
  Photomultiplier Tube},'' {\em Nucl. Instrum. Meth.}, vol.~A889, pp.~69--77,
  2018.

\bibitem{LAPPD:2016yng}
B.~W. Adams {\em et~al.}, ``{A Brief Technical History of the Large-Area
  Picosecond Photodetector (LAPPD) Collaboration},'' 3 2016.

\bibitem{Anderson09thedevelopment}
J.~Anderson {\em et~al.}, ``The development of large-area fast
  photo-detectors,'' 2009.

\bibitem{testbeam}
E.~Angelico, {\em Development of Large-Area MCP-PMT Photo-Detectors for a
  Precision Time-Of-Flight System at the Fermilab Test Beam Facility}.
\newblock PhD thesis, University of Chicago, 2020.

\bibitem{Angelico:2020xzt}
E.~Angelico, A.~Elagin, H.~J. Frisch, E.~Spieglan, B.~W. Adams, M.~R. Foley,
  and M.~J. Minot, ``{Air-Transfer Production Method for Large-Area Picosecond
  Photodetectors},'' {\em Rev. Sci. Instrum.}, vol.~91, no.~5, p.~053105, 2020.

\bibitem{Onken:2020pnv}
D.~R. Onken, F.~Moretti, J.~Caravaca, M.~Yeh, G.~D. Orebi~Gann, and E.~D.
  Bourret, ``Time response of water-based liquid scintillator from x-ray
  excitation,'' {\em Mater. Adv.}, vol.~1, pp.~71--76, 2020.

\bibitem{Caravaca:2020lfs}
J.~Caravaca, B.~J. Land, M.~Yeh, and G.~D. Orebi~Gann, ``{Characterization of
  water-based liquid scintillator for Cherenkov and scintillation
  separation},'' {\em Eur. Phys. J. C}, vol.~80, no.~9, p.~867, 2020.

\bibitem{Kaptanoglu:2019gtg}
T.~Kaptanoglu, M.~Luo, B.~Land, A.~Bacon, and J.~Klein, ``{Spectral Photon
  Sorting For Large-Scale Cherenkov and Scintillation Detectors},'' {\em Phys.
  Rev. D}, vol.~101, no.~7, p.~072002, 2020.

\bibitem{Caravaca:2016ryf}
J.~Caravaca, F.~B. Descamps, B.~J. Land, J.~Wallig, M.~Yeh, and G.~D.
  Orebi~Gann, ``{Experiment to demonstrate separation of Cherenkov and
  scintillation signals},'' {\em Phys. Rev. C}, vol.~95, no.~5, p.~055801,
  2017.

\bibitem{Gruszko:2018gzr}
J.~Gruszko, B.~Naranjo, B.~Daniel, A.~Elagin, D.~Gooding, C.~Grant, J.~Ouellet,
  and L.~Winslow, ``{Detecting Cherenkov light from 1\textendash{}2 MeV
  electrons in linear alkylbenzene},'' {\em JINST}, vol.~14, no.~02, p.~P02005,
  2019.

\bibitem{nudot}
J.~{Gruszko}, ``{NuDot: Double-Beta Decay with Direction Reconstruction in
  Liquid Scintillator},'' in {\em XXVIII International Conference on Neutrino
  Physics and Astrophysics}, p.~481, June 2018.

\bibitem{Back:2017kfo}
A.~R. Back {\em et~al.}, ``{Accelerator Neutrino Neutron Interaction Experiment
  (ANNIE): Preliminary Results and Physics Phase Proposal},'' {\em
  arXiv:1707.08222 [physics.ins-det]}, 2017.

\bibitem{Askins:2015bmb}
M.~Askins {\em et~al.}, ``{The Physics and Nuclear Nonproliferation Goals of
  WATCHMAN: A WAter CHerenkov Monitor for ANtineutrinos},'' 2 2015.

\bibitem{Askins:2019oqj}
M.~Askins {\em et~al.}, ``{Theia: An advanced optical neutrino detector},''
  {\em arXiv:1911.03501 [physics.ins-det]}, 2019.

\bibitem{Lyashenko:2019tdj}
A.~V. Lyashenko {\em et~al.}, ``{Performance of Large Area Picosecond
  Photo-Detectors (LAPPD$^{TM}$)},'' {\em Nucl. Instrum. Meth. A}, vol.~958,
  p.~162834, 2020.

\bibitem{osti_1564252}
M.~Minot {\em et~al.}, ``Large area picosecond photodetector (lappd tm ) -
  pilot production and development status,'' {\em Nuclear Instruments and
  Methods in Physics Research. Section A, Accelerators, Spectrometers,
  Detectors and Associated Equipment}, vol.~936, 12 2018.

\bibitem{Malace:2021hma}
S.~P. Malace and S.~Wood, ``{Single photoelectron identification with Incom
  LAPPD 38},'' {\em JINST}, vol.~16, no.~08, p.~P08005, 2021.

\bibitem{incom_communication}
Incom, ``{Measurement and Test Report for LAPPD 93},'' {\em Private
  Communication}, 2021.

\bibitem{Jocher:2018yhc}
G.~R. Jocher, M.~J. Wetstein, B.~Adams, K.~Nishimura, and S.~M. Usman,
  ``{Multiple-photon disambiguation on stripline-anode Micro-Channel Plates},''
  {\em Nucl. Instrum. Meth. A}, vol.~822, p.~25, 2016.

\bibitem{Dunger:2019dfo}
J.~Dunger and S.~D. Biller, ``{Multi-site Event Discrimination in Large Liquid
  Scintillation Detectors},'' {\em Nucl. Instrum. Meth. A}, vol.~943,
  p.~162420, 2019.

\bibitem{Tiras:2019ozv}
E.~Tiras, ``{Detector R\&D for ANNIE and Future Neutrino Experiments},'' in
  {\em {Meeting of the Division of Particles and Fields of the American
  Physical Society}}, 10 2019.

\bibitem{Land:2020oiz}
B.~J. Land, Z.~Bagdasarian, J.~Caravaca, M.~Smiley, M.~Yeh, and G.~D.
  Orebi~Gann, ``{MeV-scale performance of water-based and pure liquid
  scintillator detectors},'' {\em Phys. Rev. D}, vol.~103, no.~5, p.~052004,
  2021.

\bibitem{Abusleme:2020bbm}
A.~Abusleme {\em et~al.}, ``{Optimization of the JUNO liquid scintillator
  composition using a Daya Bay antineutrino detector},'' {\em Nucl. Instrum.
  Meth. A}, vol.~988, p.~164823, 2021.

\bibitem{Anderson:2020xxb}
M.~R. Anderson {\em et~al.}, ``{Development, characterisation, and deployment
  of the SNO+ liquid scintillator},'' 11 2020.

\bibitem{OKeeffe:2011dex}
H.~M. O'Keeffe, E.~O'Sullivan, and M.~C. Chen, ``{Scintillation decay time and
  pulse shape discrimination in oxygenated and deoxygenated solutions of linear
  alkylbenzene for the SNO+ experiment},'' {\em Nucl. Instrum. Meth.},
  vol.~A640, pp.~119--122, 2011.

\bibitem{spectrum_techniques}
S.~Techniques, ``{Beta Disk Sources}.''
  \url{https://www.spectrumtechniques.com/products/sources/disk-sources-and-source-sets/},
  2021.
\newblock [Accessed Sep. 17, 2021].

\bibitem{eljin}
E.~Technology, ``{Silicone Grease EJ-550}.''
  \url{https://eljentechnology.com/products/accessories/ej-550-ej-552}, 2021.
\newblock [Accessed Sep. 17, 2021].

\bibitem{r11934}
H.~Photonics, ``{Hamamatsu R11934 Datasheet}.''
  \url{https://www.hamamatsu.com/resources/pdf/etd/R11265U_H11934_TPMH1336E.pdf},
  2019.
\newblock [Accessed April 27, 2021].

\bibitem{caen}
CAEN, ``{CAEN V1742}.'' \url{https://www.caen.it/products/v1742/}, 2021.
\newblock [Accessed Sep. 17, 2021].

\bibitem{ratpac}
S.~Seibert {\em et~al.}, ``{RAT User Guide}.''
  \url{https://rat.readthedocs.io/en/latest/}, 2014.
\newblock [Accessed Oct. 29, 2019].

\bibitem{GEANT4:2002zbu}
S.~Agostinelli {\em et~al.}, ``{GEANT4--a simulation toolkit},'' {\em Nucl.
  Instrum. Meth. A}, vol.~506, pp.~250--303, 2003.

\bibitem{Lombardi:2013nla}
P.~Lombardi, F.~Ortica, G.~Ranucci, and A.~Romani, ``{Decay time and pulse
  shape discrimination of liquid scintillators based on novel solvents},'' {\em
  Nucl. Instrum. Meth.}, vol.~A701, pp.~133--144, 2013.

\bibitem{MarrodanUndagoitia:2009kq}
T.~Marrodan~Undagoitia, F.~von Feilitzsch, L.~Oberauer, W.~Potzel, A.~Ulrich,
  J.~Winter, and M.~Wurm, ``{Fluorescence decay-time constants in organic
  liquid scintillators},'' {\em Rev. Sci. Instrum.}, vol.~80, p.~043301, 2009.

\bibitem{Li2011}
X.-B. Li, H.-L. Xiao, J.~Cao, J.~Li, X.-C. Ruan, and Y.-K. Heng, ``Timing
  properties and pulse shape discrimination of {LAB}-based liquid
  scintillator,'' {\em Chinese Physics C}, vol.~35, pp.~1026--1032, nov 2011.

\bibitem{birks}
J.~B. Birks, {\em The Theory and Practice of Scintillation Counting}.
\newblock International Series of Monographs in Electronics and
  Instrumentation, Pergamon, 1964.

\bibitem{BOREXINO:2020aww}
M.~Agostini {\em et~al.}, ``{Experimental evidence of neutrinos produced in the
  CNO fusion cycle in the Sun},'' {\em Nature}, vol.~587, pp.~577--582, 2020.

\end{thebibliography}

\end{document}